\documentclass[extra] {gji}

\pdfoutput=1

\usepackage{timet}
\usepackage{graphicx}
\usepackage{lscape}
\usepackage{xcolor}
\usepackage{amsmath}
\usepackage{amssymb}
\usepackage{calrsfs}
\usepackage[mathscr]{euscript}
\usepackage{bm}

 \usepackage{lineno}

\renewcommand{\cite}{\citet}

\title[Inference of core surface magnetic and velocity fields changes]
	{Assimilation of ground and satellite magnetic measurements: inference of core surface magnetic and velocity field changes}
\author[Barrois et al.]
  {O. Barrois$^1$, M.D. Hammer$^2$, C.C. Finlay$^2$, Y. Martin$^1$ \& N. Gillet$^1$\\
  $^1$ Univ. Grenoble Alpes, CNRS, ISTerre, CS 40700 F-38058 Grenoble cedex 9, France.\\
  $^2$ Division of Geomagnetism, DTU Space, Technical University of Denmark, Lyngby DK-2800, Denmark.}
\date{Received XXX; in original form XXX}
\pagerange{\pageref{firstpage}--\pageref{lastpage}}
\volume{XXX}
\pubyear{2018}

\begin{document}

\label{firstpage}

\maketitle

\begin{summary}

We jointly invert for magnetic and velocity fields at the core surface over the period 1997--2017, directly using ground-based observatory time series and measurements from the CHAMP and \textit{Swarm} satellites.
Satellite data are reduced to the form of virtual observatory time series distributed on a regular grid in space.
Such a sequential storage helps incorporate voluminous modern magnetic data into a stochastic Kalman filter, whereby spatial constraints are incorporated based on a norm derived from statistics of a numerical geodynamo model.
Our algorithm produces consistent solutions both in terms of the misfit to the data and the estimated posterior model uncertainties.  
We retrieve core flow features previously documented from the analysis of spherical harmonic field models, such as the eccentric anti-cyclonic gyre. 
We find enhanced diffusion patterns under both Indonesia and Africa.
In contrast to a steady flow that is strong under the Atlantic hemisphere but very weak below the Pacific, interannual motions appear evenly distributed over the two hemispheres.
Recovered interannual to decadal flow changes are predominantly symmetrical with respect to the equator outside the tangent cylinder. 
In contrast, under the Northern Pacific we find an intensification of a high latitude jet, but see no evidence for a corresponding feature in the Southern hemisphere. 
The largest flow accelerations that we isolate over the studied era are associated with meanders, attached to the equatorward meridional branch of the planetary gyre in the Eastern hemisphere, that is linked to the appearance of an eastward equatorial jet below the Western Pacific.

\end{summary}


\section{Introduction}
\label{sec:state_of_art}

Inferring information on the motions of the liquid outer core of the Earth requires properly separating the numerous sources of observed magnetic fields (geodynamo, crustal magnetisation, ionospheric and magnetospheric currents and their Earth induced counterparts).
To circumvent some of the leakage issues, magnetic field models are often built using regularizations, to ensure spectral convergence of the core field and its time variations. 
This prevents a proper assessment of a posteriori errors on model coefficients.
When these are used as data in reconstructions of the core dynamics, it can lead to biased estimates.
Furthermore, by proceeding in successive steps (to a field model and then on to the core flow), one loses information.

From the early 1990's alternative avenues of research arose, through which field models were built under topological constraints derived from physical insights. 
\citet{constable1993geomagnetic} and  \cite{o1997frozen} proposed algorithms to apply, on single epoch pairs of models, magnetic flux conservation conditions at the core-mantle boundary (CMB) that are appropriate assuming that magnetic diffusion is negligible. 
Along the same lines, \citet{jackson2007models} added a constraint on the radial vorticity.
They showed that it was possible for a magnetic model to satisfy both these topological conditions, and the constraint from magnetic observations, from the late 19th century onwards.

Conversely, \citet{chulliat2010observation} tested the validity of the frozen flux hypothesis using data from Magsat, Oersted and CHAMP satellite missions.
They found an increase of the data misfit in some areas, potentially suggesting local failures of the constraint.
Such studies motivated the co-estimation, from magnetic observations, of both the field and the flow, imposing with a weak formalism the frozen flux radial induction equation at the CMB \citep{lesur2010modelling,wardinski2012extended}. 
They concluded that the frozen flux constraint remained compatible with ground-based and satellite magnetic records.
Pursuing an alternative approach, \citet{beggan2009forecasting} and \citet{whaler2015derivation} obtained piecewise constant or linear flow models directly from magnetic data \citep[see also][]{whaler2016decadal}.

One limitation though of such approaches is related to the uncertainties associated with the large scale induction equation itself (and associated null-flux curves), assuming models truncated at spherical harmonic degree $n\simeq13$ \citep{gillet2009ensemble}. 
Subgrid-scale effects arising due to the non-linear induction process \citep[e.g.][]{eymin2005core,pais2008quasi,gillet2009ensemble,baerenzung2016flow} turn out to be the main source of uncertainty in the recovery of core surface flows from modern geomagnetic records.
\cite{barrois2017contributions} -- hereafter referred as BGA17 -- illustrate how ignoring their impact leads to severely biased flow models \citep[see also][on the reliability of core flow reconstructions]{baerenzung2017modeling}.

BGA17 furthermore show from the analysis of geodynamo simulations that magnetic diffusion at the core surface, enslaved to poloidal flow below the CMB, affects the recorded field changes at all time-scales including rapid changes. 
This may seem at odds with the often used assumption of negligible magnetic diffusion that follows the argument of a high magnetic Reynolds number for large-scale motions in the core  \citep[see][]{holme2015TOG}.

In the present work we invert, from magnetic field observations collected at and above the Earth's surface, for both the magnetic and velocity fields at the core surface, taking into account both magnetic diffusion and subgrid induction.
We merge spatial information provided by numerical simulations, specifically from the Coupled Earth dynamo (CED) model \citep{aubert2013bottom} and temporal constraints coming from a restriction of the field evolution to a chosen class of stochastic process. 
The sequential algorithm of BGA17, which considers as input data time series of spherical harmonic coefficients of the main field, is extended to account for both virtual observatory \citep{mandea2006new} and ground observatory time series that cover the period 1997--2017.
Our approach has similarities with the previous works of \citet{gillet2015stochastic} and \citet{baerenzung2016flow}, which favoured flat flow spatial spectra at the CMB, since the spatial dynamo norm employed here departs from the norms often employed to ensure spectral convergence.
In addition, our stochastic framework allows us to discuss posterior model errors for both the flow and the magnetic field.

The paper is organised as follows. 
In section \S\ref{sec:Method} we describe the ground-based observatory data and satellite-based virtual observatory data, and the methodology we follow to recover magnetic and velocity fields at the CMB.
In section \S\ref{sec: mag}, we present our resulting geomagnetic model and its associated uncertainties, before we analyse in \S\ref{sec:Core_flows} our core flow solutions.
Finally, implications for our understanding of the core dynamics and possible further improvements for the algorithm are given in section \S\ref{sec:Conclusion}.

\section{Methodology}
\label{sec:Method}

\subsection{Ground-based and virtual observatory data}
\label{sec:VO_motivation}

\subsubsection{Ground observatory data}

We use magnetic measurements made at 186 ground observatories (GO) covering the period 1997--2017.
Hourly mean values are taken from the BGS database\footnote{{\tt ftp://ftp.nerc-murchison.ac.uk/geomag/Swarm/AUX\_OBS}}, version 0111, using Intermagnet and WDC Edinburgh data as available in May 2017. 
The data have been checked and corrected for known baseline jumps \citep{macmillan2013observatory}. 
`Revised monthly means' were then derived from these hourly means, following the procedure described by \citet{olsen2014chaos}.  Briefly, predictions of the large-scale magnetospheric field (and the associated induced field) from the CHAOS-6 field model, as well as predictions for the ionospheric Sq field (and the associated induced field) from the CM4 model \citep{sabaka2004extending} are subtracted from the hourly mean values, and then robust (Huber-weighted) monthly mean values are computed using an iterative-reweighting procedure.
Annual differences of such revised monthly means are routinely used in deriving the CHAOS series of field models and in order to study high resolution secular variation since, compared with simple monthly means, they are less contaminated by external field effects.
Here, since we also wish to use the field itself for model construction, the median difference between each series and CHAOS-6 predictions was removed, in order to account in a simple way for the bias due to unmodelled crustal fields.
In order to obtain the same sampling rate as that adopted for the virtual observatory series described below, the revised monthly mean series were finally averaged over 4 months windows to obtain the GO series used in our data assimilation scheme.

\subsubsection{Virtual observatory data}

In addition to GO data, we make use of satellite measurements from the CHAMP and \textit{Swarm} missions covering respectively 2000--2010 and  2014--2017, through so-called virtual observatory (VO) data  \citep{mandea2006new, olsen2007investigation}. 
These provide a regular spatial and temporal sampling of the global field, convenient for our Kalman filter algorithm (detailed in \S\ref{sec:algo_modif}) and involve estimates from an easily manageable number of locations, which has computational advantages.

VO data were computed using measurements collected by the CHAMP vector field magnetometer between July 2000 and September 2010 and from the \textit{Swarm} vector field magnetometers, onboard all three satellites (\textit{Alpha}, \textit{Bravo}, \textit{Charlie}), between January 2014 and April 2017. 
Starting from the CHAMP MAG-L3 and \textit{Swarm} Level 1b MAG-L, version 0501, data products, we sub-sampled at 15s intervals the data in the vector field magnetometer (VFM) frame. 
Using the Euler rotation angles as given by the CHAOS-6-x3 model (which was based on \textit{Swarm} and ground observation data up until April 2017\footnote{\tt http://www.spacecenter.dk/files/magnetic-models/CHAOS-6/}), we rotated the VFM data into an Earth-Centered Earth-Fixed (ECEF) coordinate frame.

Measurements from known problematic days were removed, for instance where satellite manoeuvres happened. 
Furthermore, gross data outliers with deviations more than 500 nT from CHAOS-6-x3 field model predictions were rejected.
Based on previous studies of VO estimates \citep[e.g.,][]{beggan2009biased}, we then employed data selection criteria retaining only data for which: 
\begin{enumerate}
\item[-] the sun was at maximum $10^{\circ}$ above horizon;
\item[-] geomagnetic activity index $\mathrm{K_p}<3^{\mathrm{o}}$;
\item[-] the $RC$ disturbance index \citep{olsen2014chaos} had $\vert d\mathrm{RC}/dt \vert < 3$ nT/h;
\item[-] merging electric field at the magnetopause $\mathrm{E_m}\leq 0.8$ mV/m, with $E_m=0.33 v^{4/3} B_t^{2/3} \mathrm{sin}(\vert \Theta \vert/2)$. $v$ is the solar wind speed, $\Theta=\mathrm{arctan}(B_y/B_z)$ and $B_t=\sqrt{B_y^2+B_z^2}$. 
$B_y$ and $B_z$ are components of the interplanetary magnetic field (IMF) in the geocentric solar magnetospheric (GSM) coordinate system, calculated using 2 hourly means of 1-min values of the IMF and solar wind extracted from the OMNI database\footnote{\tt http://omniweb.gsfc.nasa.gov}; 
\item[-]  IMF $B_z>0$ nT and IMF $ \vert B_y \vert<10$ nT, again based on 2 hourly mean of 1 minute values. 
\end{enumerate}

Following this data selection, estimates of the fields due to various unmodelled sources were next removed from the data: 
\begin{enumerate}
\item[(i)] the magnetospheric and its induced fields as given by the CHAOS-6-x3 model;
\item[(ii)] the ionospheric and its induced fields as given by the CM4 model \citep{sabaka2004extending};
\item[(iii)] the static internal field for spherical harmonic degrees $n>20$ given by the CHAOS-6-x3 model.
\end{enumerate}
Although imperfect, in our opinion it is more consistent to remove such estimates rather to ignore known field sources.

Based on this data we then carried robust inversions for time-averaged point estimates (i.e. VOs) using data windows of 4 months width (60 days each side of an epoch $t_j$).  In order to aid the robust inversion procedure in identifying and downweighting outliers,  following  \citet{olsen2007investigation} as a pre-processing step, we  also removed a time-dependent internal field, here taken from the CHAOS-6-x3 model \citep{finlay2016recent}, for spherical harmonic degrees 1 to 20, within each four month window. The CHAOS-6x-3 prediction at the target point and time was then added back at the end of the analysis. 
Note that this does not prevent our 4-monthly VO series, and the derived SV series from departing from CHAOS-6x-3; information about the time-dependence within each 4 month window is however lost.    

We assume that the residual field $\tilde{\bf B}$, after the removal of the time-dependent internal field from the CHAOS-6-x3, can be represented as the gradient of a scalar potential $V$, i.e. 
\begin{linenomath*}\begin{eqnarray}
\tilde{\bf B}  = -\nabla V\,.
\label{eq:V_VO_grad}
\end{eqnarray}\end{linenomath*}
The residual field and associated positions are transformed into a local Cartesian coordinate system with origin at the VO points of interest, with $x$ pointing towards geographic South, $y$ pointing towards East and $z$ pointing upwards. 
We use an expansion of the local potential up to cubic terms. 
Because the geomagnetic field is irrotational ($\nabla \times \tilde{\bf B}=0$) and solenoidal ($\nabla\cdot\tilde{\bf B}=0$), this local potential is entirely determined by 15 independent parameters:
\begin{linenomath*}\begin{eqnarray}
\label{eq:V_VO_param}
V(x,y,z) = v_{x}x + v_{y}y + v_{z}z + v_{xx}x^2 + v_{yy}y^2 - (v_{xx} + v_{yy})z^{2}\\
\nonumber
+ 2v_{xy}xy + 2v_{xz}xz + 2v_{yz}yz - (v_{xyy} + v_{xzz})x^3\\
\nonumber 
+ 3v_{xxy}x^2y + 3v_{xxz}x^2z + 3v_{xyy}xy^2 + 3v_{xzz}xz^2 + 6v_{xyz}xyz\\
\nonumber  
- (v_{xxy} - v_{yzz})y^3 + 3v_{yyz}y^2z + 3v_{yzz}yz^2 - (v_{xxz} + v_{yyz})z^3\,.
\end{eqnarray}\end{linenomath*}

For each VO position vector ${\bf r}_{k}=(\theta_k,\phi_k,r_k)$ and at epoch $t_j$, all data positioned within a cylinder of radius 850km ($\approx 7.5^{\circ}$) of the VO target ${\bf r}_{k}$, and within 60 days either side of  $t_j$ were used to build a local data vector ${\bf d}^{k,j}$.   These data are then related to the 15 parameters defining the VO potential model ${\bf m}_{vo}^{k,j}$ at that site and epoch via ${\bf d}^{k,j}=  \underline{\underline{{\bf g}}}^{k,j} 
{\bf m}_{vo}^{k,j}$, where the elements of the matrix $ \underline{\underline{{\bf g}}}^{k,j}$ are determined from (\ref{eq:V_VO_grad}) and (\ref{eq:V_VO_param}).

Rather than working directly with ${\bf d}^{k,j}$ in deriving ${\bf m}_{vo}^{k,j}$  we make use of along-track and East-West (using  \textit{Swarm} \textit{Alpha} and \textit{Charlie} only) sums and differences of the magnetic field components.  An advantage of using field differences is that these have a reduced sensitivity to large-scale external signals, although data sums also need to be included in order to ensure sufficient information on the longer wavelengths core field.  Using sums and differences has been found advantageous in a number of other field modelling efforts \citep{sabaka2015cm5,olsen2015swarm}. We calculate along-track (AT) sums ($\Sigma$) and differences ($\Delta$) as
\begin{linenomath*}\begin{eqnarray}
\label{eq:AT data}
\left\{
\begin{array}{rl}
\Sigma d_i^{AT} = &[\tilde{B}_i(\mathbf{r},t) + \tilde{B}_i(\mathbf{r}+\delta \mathbf{r},t+15\mathrm{s})]/2\\
\Delta d_i^{\mathrm{AT}} = &[\tilde{B}_i(\mathbf{r},t) - \tilde{B}_i(\mathbf{r}+\delta \mathbf{r},t+15\mathrm{s})]
\end{array}
\right.\,.
\end{eqnarray}\end{linenomath*}
$\tilde{B}_i={\mathbf{1}_i}\cdot\mathbf{\tilde{B}}(\mathbf{r})$ are the residual magnetic field components in spherical polar coordinates (where $i=r,\theta$ or $\phi$, and ${\mathbf{1}_i}$ are unit vectors). 
The East-West cross-track (CT) sums and differences between are calculated as
\begin{linenomath*}\begin{eqnarray}
\label{eq:CT data}
\left\{
\begin{array}{rl}
\Sigma d_i^{\mathrm{CT}} = &[\tilde{B}_i^{\mathrm{Alpha}}(\mathbf{r}_1,t_1) + \tilde{B}_i^{\mathrm{Charlie}}(\mathbf{r}_2,t_2)]/2\\ 
\Delta d_i^{\mathrm{CT}} = &[\tilde{B}_i^{\mathrm{Alpha}}(\mathbf{r}_1,t_1) - \tilde{B}_i^{\mathrm{Charlie}}(\mathbf{r}_2,t_2)]
\end{array}
\right.\,.
\end{eqnarray}\end{linenomath*}
Here, for a given orbit of \textit{Alpha} we select the corresponding \textit{Charlie} data to be the one closest in colatitude such that $\vert\delta t\vert=\vert t_1-t_2\vert<50s$.  Crucially, in order to relate these sums and differences to the VO model parameters, we also take sums and differences of the elements of the design matrices $ \underline{\underline{{\bf g}}}^{k,j}$ associated with the predictions of the VO model for the field components at the individual data locations.  This results in a design matrix  
 \begin{linenomath*}\begin{equation}
 \underline{\underline{{\bf G}}}^{k,j} = \begin{bmatrix}   \underline{\underline{\Sigma {\bf g}}}^{k,j} \\  \underline{\underline{\Delta {\bf g}}}^{k,j} \end{bmatrix} 
 \label{eq:design_VO}
\end{equation}\end{linenomath*}
 associated with the data vector $ {\bf D}^{k,j} = \left[ \Sigma {\bf d}^{k,j} \, \,\Delta {\bf d}^{k,j} \right]^T$.  In this way we fully account for the change in the unit vectors associated with the two locations contributing to the sums and differences when deriving the parameters 
 ${\bf m}_{vo}^{k,j}$.  The inversion for each  ${\bf m}_{vo}^{k,j}$ is carried out via a robust Huber weighted least-square fit
\begin{linenomath*}\begin{eqnarray}
 {\bf m}_{vo}^{k,j} = \left[( \underline{\underline{{\bf G}}}^{k,j})^T {\bf W} \underline{\underline{{\bf G}}}^{k,j}\right]^{-1} ( \underline{\underline{{\bf G}}}^{k,j})^T  {\bf D}^{k,j}
\end{eqnarray}\end{linenomath*}
where ${\bf W}$ is a diagonal vector of Huber weights that ensures a robust solution \citep{olsen2002model,sabaka2004extending} and are iteratively updated until convergence. Once  ${\bf m}_{vo}^{k,j}$ is determined,  the three field components at the site and epoch of interest, $\tilde{\bf B}_k({\bf r}_{k}, t_j) = - \nabla V_k({\bf r}_{k},t_j)$, are computed and added back on to the CHAOS-6-x3 prediction for the internal field (for degrees 1-14 only, to avoid as far as possible the lithospheric field) at the target location. 

We constructed VO estimates at $P_{\mathrm{VO}}=200$ locations, with a spacing of about 1600 km ($\approx 14^{\circ}$, see dots in Figure \ref{fig:VO+GO_loc}), located in an approximately equal area grid based on the spherical surface partition algorithm of \citet{leopardi2006partition}.  The altitude of the VOs are 300km and 500km during the CHAMP and \textit{Swarm} periods, respectively.  Using predictions of the three components ($B_r$, $B_\theta$, $B_\phi$) of the magnetic field at $P_{\mathrm{VO}}$ locations, we finally obtain $3P_{\mathrm{VO}}$ time series (i.e. one point every 4 months during CHAMP and Swarm times, 48 epochs in all), stored in a vector ${\bf y}_{\mathrm{VO}}(t)$. The SV was computed as annual differences of the 4 month time series.

\subsubsection{Uncertainty estimates for the GO and VO series}

In order to obtain as much information as possible from the GO and VO data, while at the same time seeking to avoid overfitting them, it is important that appropriate uncertainty estimates are specified for each time series.
We define ${\bf C}_{\mathrm{GO}}$ and ${\bf C}_{\mathrm{VO}}$ to be the measurement error cross-covariance matrices for GO and VO data  at each epoch, of sizes respectively $3P_{GO} \times 3P_{GO}$ and $3P_{VO} \times 3P_{VO}$.
Data errors are supposed independent of time.
Different data uncertainties are assigned for the VO's derived from CHAMP and {\it Swarm} respectively.

Regarding the GO time series described above, we follow a similar approach to that used in CHAOS field model series \citep{olsen2014chaos, finlay2016recent} and derive uncertainty estimates as follows. 
A three-by three covariance matrix was computed for each observatory location from the time-series of the three components, after removing the predictions of the CHAOS-6 field model and de-trending.
The square root of the diagonal elements of these covariance matrices were taken to be the uncertainty estimates for each component at each observatory.  
The same procedure was applied to both the MF and SV series.

\begin{figure*}
\centerline{
	\includegraphics[width=.7\linewidth]{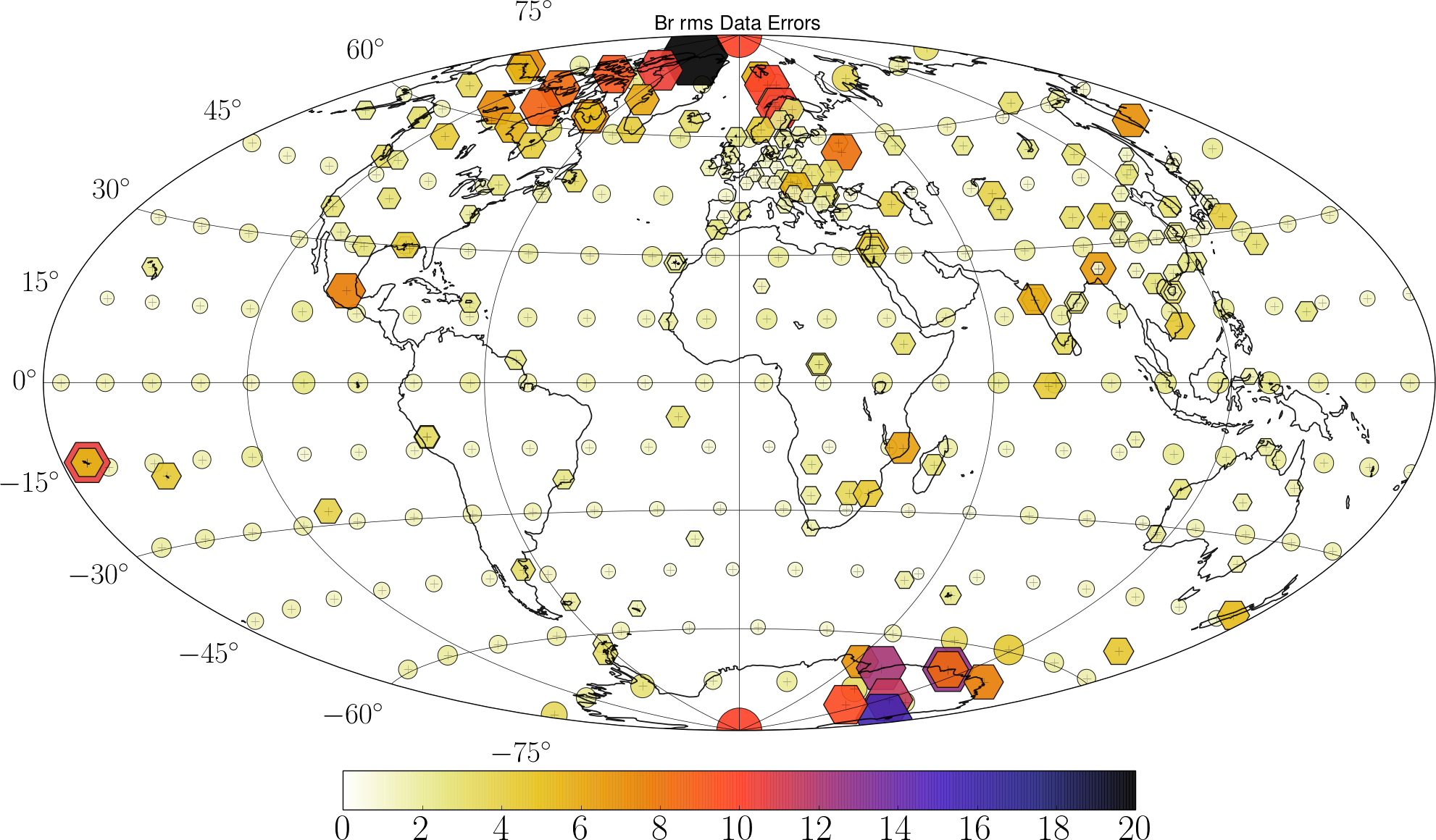}}
\centerline{
	\includegraphics[width=.7\linewidth]{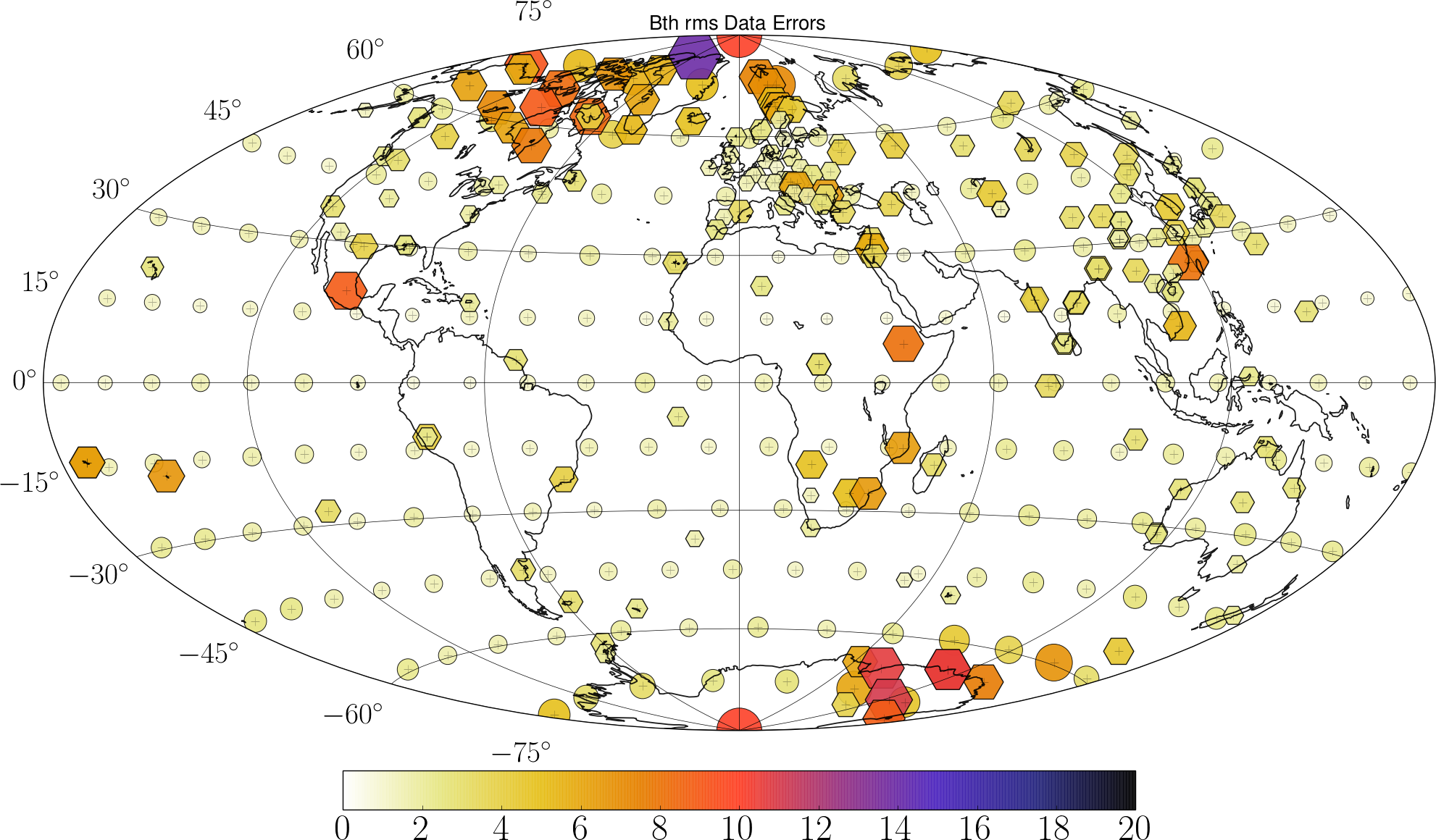}}
\centerline{
	\includegraphics[width=.7\linewidth]{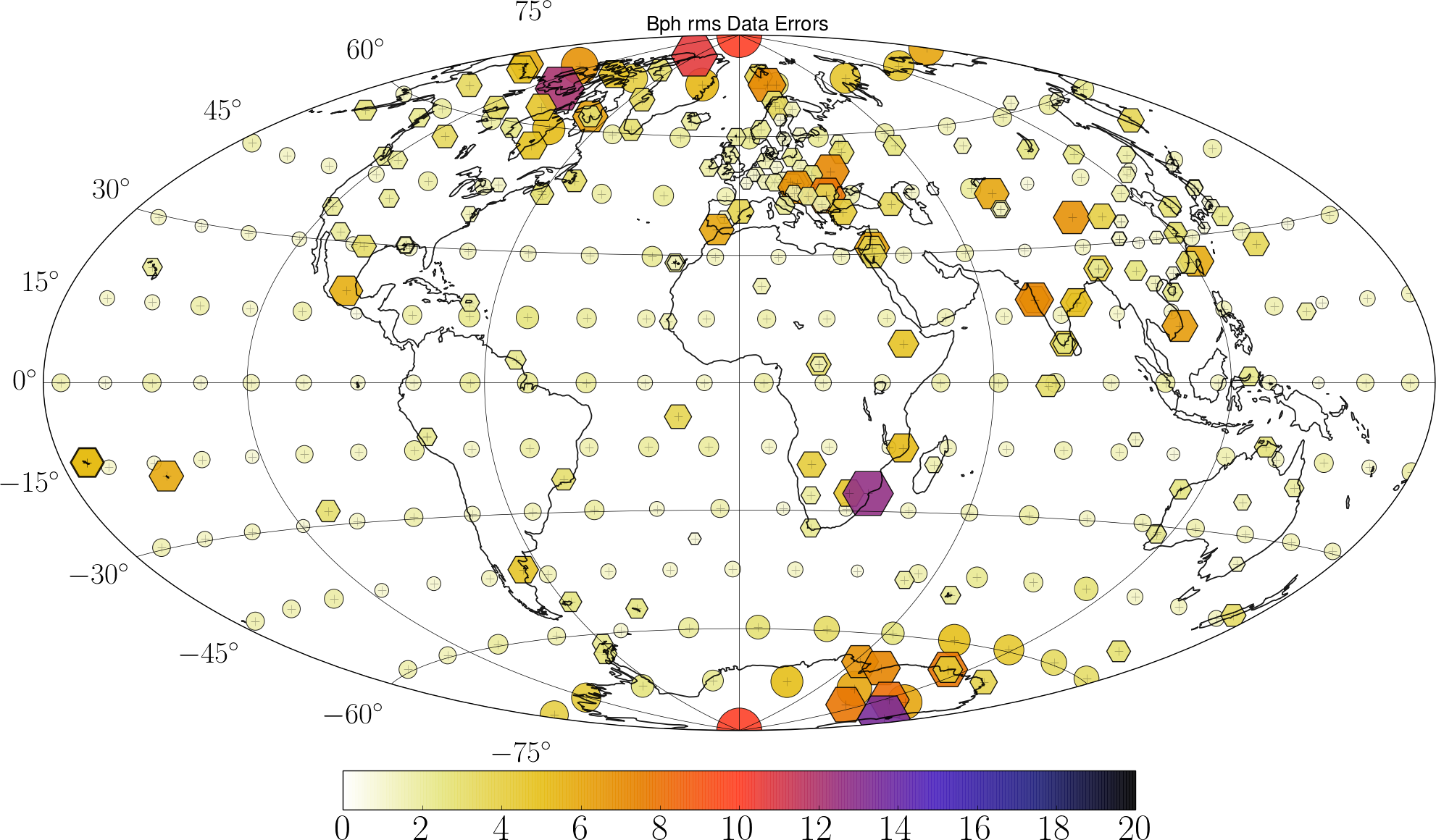}}
	\caption{
SV observation error estimates (colorscale in $nT/y$) at all location where GOs (hexagons) and VOs (circles) are used in this study, for the 3 components $\dot{B}_r$, $\dot{B}_\theta$ and $\dot{B}_\phi$ (from top to bottom).
The size of the markers is proportional to the magnitude of the \textit{a-priori} error estimates.
	\label{fig:VO+GO_loc}
	}
\end{figure*}

For consistency, a very similar procedure was also applied to the VO series in order to obtain their uncertainty estimates.
For each VO location, covariances were calculated between the time series of the three components (after removing from each series the predictions of the CHAOS-6 model and de-trending), in order to obtain a three-by three covariance matrix.
A robust procedure for calculating the covariances \citep[using the Minimum Covariance Determinant estimator,][]{verboven2005libra} was employed.
However, only the square-root of the diagonal elements of the covariance matrices were taken to be the uncertainty estimates for each series, with similar procedures applied to both MF and SV series.
To illustrate the range of the adopted uncertainty estimates, we show in Figure \ref{fig:VO+GO_loc} the r.m.s. SV uncertainty estimates for all locations where data (GO or VO) are used in this study.

Note that by using only the diagonal elements of ${\bf C}_{\mathrm{GO}}$ and ${\bf C}_{\mathrm{VO}}$ we effectively consider the errors on each GO and VO series to be uncorrelated with the errors on other series. 
In reality errors between components and between series will be correlated. 
This can be taken into account using full (i.e. dense) covariance matrices. 
It is however challenging to estimate cross-covariances for matrices of size larger than the length of the contributing times series (consisting of one sample every four months).
We therefore postpone this step to future studies. 
Instead, by restricting to only 200 VO locations and ensuring that there was little overlap between the VO search radii we reduce as far as possible the correlations between distinct VO series.

Finally, we concatenate the above GO and VO main field data vectors for each epoch into ${\bf y}^o(t)=[{\bf y}_{\mathrm{VO}}^T\, {\bf y}_{\mathrm{GO}}^T]^T$.
The associated observation errors covariance matrix ${\bf R}_{yy}$, of rank $P=3 P_{\mathrm{VO}}+ 3 P_{\mathrm{GO}}$, is thus derived from the diagonals of ${\bf C}_{\mathrm{VO}}$ and ${\bf C}_{\mathrm{GO}}$.
In the next section we will consider both main field and secular variation data.
SV data $\dot{\bf y}^o(t)$ are computed as annual differences of the four monthly (GO or VO) series.
We follow the same approach as above to estimate the SV data errors variances (shown in Figure \ref{fig:VO+GO_loc}) that are stored in a diagonal matrix ${\bf R}_{\dot{y}\dot{y}}$ of rank $P$.

\subsection{Re-analysis of GO and VO data ground and satellite magnetic observations}
\label{sec:algo_modif}

The assimilation algorithm used in the present study is essentially the one derived by BGA17 (see their table 2 for a summary). 
It is a sequential tool, consisting of a succession of forecast and analysis steps. 
The main modifications concern the direct integration of observations at and above the Earth's surface, while BGA17 considered data in the form of MF and SV spherical harmonic coefficients.
We begin by recalling the main points of our stochastic forecast model, before we go on to describe the changes implemented in the present study regarding the analysis step. 
These essentially concern the observation operator linking the state variables to the observations.

\subsubsection{Stochastic forecast model}
\label{sec: forecast}

We forecast the evolution of the radial magnetic field, $B_r$, at the CMB using the radial component of the induction equation, written as
 \begin{linenomath*}\begin{eqnarray}
\label{eq: ind eqn}
\frac{\partial \overline{B}_r}{\partial t} = -\overline{\nabla_h\cdot\left({\bm u}_H\overline{B}_r\right)} + e_r + d_r({\bm u}_H,\overline{B}_r) \,,
\end{eqnarray}\end{linenomath*}
where overlines mean the projection onto large length-scales. 
$e_r$ stands for the subgrid induction processes arising due to the unresolved magnetic field at small length-scales, ${\bm u}_H$ is the horizontal flow, and $d_r$, enslaved to $\overline{B}_r$ and ${\bm u}_H$, approximates the radial component of the diffusion operator (see below). 
The evolutions of $e_r$ and ${\bm u}_H$ are governed by order one auto-regressive stochastic processes,
 \begin{linenomath*}\begin{eqnarray}
\label{eq: subgrid}
\frac{\mathrm{d} e_r}{\mathrm{d} t} + \frac{e_r}{\tau_e}=\zeta_e \,,
\end{eqnarray}\end{linenomath*}
 \begin{linenomath*}\begin{eqnarray}
\label{eq: flow}
\frac{\mathrm{d} {\bm u}_H}{\mathrm{d} t} + \frac{({\bm u}_H - \hat{\bm u}_H)}{\tau_u}=\zeta_u \,,
\end{eqnarray}\end{linenomath*}
with $\zeta_e$ and $\zeta_u$ white noise processes, and $\hat{\bm u}_H$ the background flow model (obtained as the time-averaged flow from the CED model). 
These processes come from the same family of process as employed by \cite{baerenzung2017modeling}.
For each process, an effective restoring force is implemented via single time scales that we respectively fix as $\tau_e=10$ yrs and $\tau_u=30$ yrs.
Spatial cross-covariances of the two above fields are derived from statistics of a free run of the CED \citep{aubert2013bottom}.

The advected fields $e_r$, ${\bm u}_H$, $\overline{B}_r$ and $d_r$ are represented through spherical harmonics, whose coefficients are stored in vectors ${\bf e}(t)$, ${\bf u}(t)$, ${\bf b}(t)$ and ${\bf d}(t)$, respectively.
Diffusion in equation (\ref{eq: ind eqn}), and its dependence on $e_r$ and ${\bf u}_H$, is also an expression of cross-covariances extracted from the CED (involving the radial magnetic field below the CMB).
The projection onto large length-scales is processed in the spectral domain, restricting the induction equation (and thus the expansion of the fields $e_r$, $\overline{B}_r$ and $d_r$) to spherical harmonic degrees $n\le n_b=14$, while the velocity field is truncated at $n_u=18$.
We write as $\dot{\bf b}(t)$ the vector of SV spherical harmonic coefficients. 

\subsubsection{Integrating ground and satellite data in the assimilation tool}
\label{sec: assim}

We write as ${\bf M}$ the operator that links the vector ${\bf b}(t)$ to the three components main field observations ${\bf y}(t)$ in the spatial domain \citep[e.g.][]{olsen2010separation}: 
\begin{linenomath*}\begin{eqnarray}
\label{eq:SH_spec2phy}
{\bf y}(t) = {\bf M}{\bf b}(t) \,.
\end{eqnarray}\end{linenomath*}
At each epoch it is of size $n_o\times n_b(n_b+2)$, with $n_o=3(P_{\mathrm{VO}}+P_{\mathrm{GO}})$ the size of the observation vector. 
The matrix ${\bf M}$ is composed of sub-matrices ${\bf M}_{r}$, ${\bf M}_{\theta}$ and ${\bf M}_{\phi}$, depending on the considered component of the magnetic field.
In practice, elements of the matrix are, for a column $j$ corresponding to a coefficient $g_{n_j}^{m_j}$, and a line $i$ to an observation at a coordinate ${\bf r}_{i} = (r_{i},\theta_{i},\phi_{i})$,
\begin{linenomath*}\begin{eqnarray}
\label{eq:Mr_spec2phy}
{\bf M}_{r i,j} &=& (n_{j} +1 ) \left( \dfrac{a^{\oplus}}{r_{i}} \right)^{n_{j} + 2} \mathscr{P}_{n}^{m}(\theta_{i}) \cos( m_{j} \phi_{i} ) \,, \\
\label{eq:Mth_spec2phy}
{\bf M}_{\theta i,j} &=& \left( \dfrac{a^{\oplus}}{r_{i}} \right)^{n_{j} + 2} \dfrac{d \mathscr{P}_{n}^{m}(\theta_{i})}{d \theta} \cos( m_{j} \phi_{i} ) \,, \\
\label{eq:Mph_spec2phy}
{\bf M}_{\phi i,j} &=& \left( \dfrac{a^{\oplus}}{r_{i}} \right)^{n_{j} + 2} \dfrac{m_{j} \mathscr{P}_{n}^{m}(\theta_{i})}{\sin( \theta_{i})} (-1)\sin( m_{j} \phi_{i} ) \,.
\end{eqnarray}\end{linenomath*}
For a line $j$ corresponding to a coefficient $h_{n_j}^{m_j}$, the function $\sin$ replaces $\cos$ in (\ref{eq:Mr_spec2phy}) and (\ref{eq:Mth_spec2phy}), and $\cos$ replaces $(-1)\sin$ in (\ref{eq:Mph_spec2phy}).
$a^{\oplus}=6371.2$ km is the Earth's spherical reference radius and $\mathscr{P}_{n}^{m}$ are the Legendre polynomials.

The analysis in the Kalman filter algorithm employed by BGA17 consists of two steps: 
first an analysis of the vector ${\bf b}$ containing MF spherical harmonic coefficients from MF spherical harmonic coefficients data, and second an analysis of the vector ${\bf z}$ (that concatenates ${\bf u}$ and ${\bf e}$) from SV spherical harmonic coefficients data. 
Writing as ${\bf P}_{bb}^f$ the forecast model covariance matrix for ${\bf b}$, the first analysis (equation (19) of BGA17) is replaced here by 
\begin{linenomath*}\begin{eqnarray}
\label{eq:ASH-KF_spec2phy_step1}
\forall k\in [1,N_m],\; \\
{\bf b}^{ka}(t_a)= {\bf b}^{kf}(t_a) +  
{\bf P}_{bb}^f{\bf M}^T\left[{\bf M} {\bf P}_{bb}^f {\bf M}^{T} + {\bf R}_{yy}\right]^{-1} \nonumber \\
\cdot \left({\bf y}^{ko}(t_a) - {\bf M}{\bf b}^{kf}(t_a)\right)\,, \nonumber
\end{eqnarray}\end{linenomath*}
with $t_a$ the analysis epoch and the superscript $k$ referring to the $k^{th}$ realization within an ensemble chosen to be of size $N_m=50$.  
Writing as ${\bf P}_{zz}^f$ the forecast model covariance matrix for ${\bf z}$, the second analysis (equation (20) of BGA17) is replaced here by 
\begin{linenomath*}\begin{eqnarray}
\label{eq:ASH-KF_spec2phy_step2}
\forall k\in [1,N_m],\; \\
{\bf z}^{ka}(t_a)= {\bf z}^{kf}(t_a) +  
{\bf P}_{zz}^f {\bf G}^{kT} \left[ {\bf G}^k {\bf P}_{zz}^f {\bf G}^{kT} + {\bf R}_{\dot{y}\dot{y}}\right]^{-1} \nonumber \\
\cdot \left(\delta\dot{\bf y}^{ko}(t_a) - {\bf G}^k{\bf z}^{kf}(t_a)\right)\,, \nonumber
\end{eqnarray}\end{linenomath*}
where the new observation operator is ${\bf G}^k={\bf M}{\bf H}({\bf b}^{ka})$, with ${\bf H}$ as defined in BGA17.
Here $\delta \dot{\bf y}^{ko}(t_a) = \dot{\bf y}^{ko}(t_a) - {\bf M}{\bf d}^{kf}(t_a)$ are the direct SV observations corrected by the forecast contribution from diffusion to the radial induction equation. 
This latter is sought iteratively at each analysis step, as in BGA17.
Note that we consider an ensemble of observations ${\bf y}^o$ and $\dot{\bf y}^o$, which are perturbed by random noise according to respectively ${\bf R}_{yy}$ and ${\bf R}_{\dot{y}\dot{y}}$. 
We recall that we consider in equations (\ref{eq:ASH-KF_spec2phy_step1}--\ref{eq:ASH-KF_spec2phy_step2}) forecast covariance matrices ${\bf P}_{zz}$ and ${\bf P}_{bb}^f$ that are frozen throughout the re-analysis period. 
These are derived directly from the CED cross-covariances on ${\bf b}$, ${\bf u}$ and ${\bf e}$ spherical harmonic coefficients, involving scaling prefactors obtained analytically from the stochastic model presented in \S\ref{sec: forecast}  (see BGA17 for details). 
For comparison, \cite{baerenzung2017modeling} employ a full implementation of the Ensemble Kalman filter \citep{evensen2003ensemble}, i.e. they update the cross-covariances at each analysis step, requiring many more realizations to obtain well-conditioned matrices. 

\begin{figure*}
\centerline{
	\includegraphics[width=.7\linewidth]{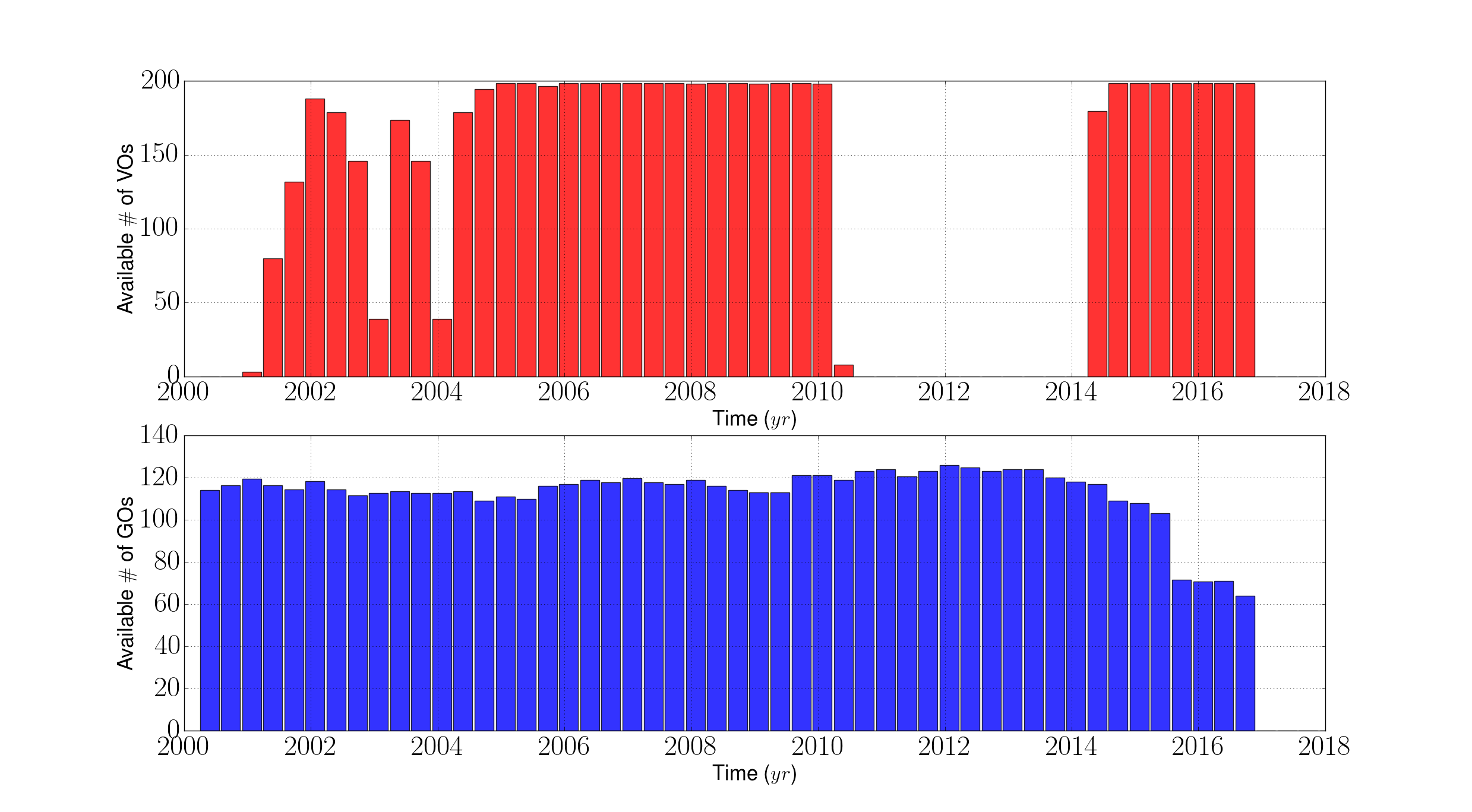}}
	\caption{
Time evolution of the number of SV data points (VOs in red, GOs in blue).
	\label{fig:VO+GO_time}
	}
\end{figure*}

Finally, an extra complexity arises because the number of observation sites changes over time.
Indeed, because of the selection criteria, the number of satellite data available may not always be sufficient to make a reliable VO estimate. 
Under these conditions the VO data point is considered to be absent: 
the associated elements of the data vector ${\bf y}^o(t)$ at a given time $t$ are removed, together with the corresponding lines and columns of ${\bf R}_{yy}$, and the corresponding lines of the matrix ${\bf M}$ (and thus ${\bf G}$).
This procedure is performed during each analysis.
Thus, the size $P$ of the data vector changes through time, reflecting the changing number of available satellite observations through time (see Figure~\ref{fig:VO+GO_time}).

To summarise, in this study we work with predictions made by spherical harmonic coefficients that are projected in physical space, where they are adjusted during the analysis step according to the observations and the covariance matrices.
As such, our algorithm is still based almost entirely on the spectral domain; only the analysis steps are performed in physical space, in order to match the observed magnetic field data.
Notice that we corrected for two mistakes in the implementation of the algorithm by BGA17: a sign error in the background flow $\hat{\bf u}$, and off-diagonal elements of the covariance matrix for ${\bf e}$ were non intentionally ignored. 
Performing comparisons between re-analyses before and after correction, we found two consequences: a reduction of the dispersion within the ensemble of realizations, and an (almost stationary) shift in the analysed diffusion for some coefficients (including the axial dipole, see \S\ref{sec:Geomag_model}). 
This latter is almost entirely compensated by a shift in the analysed $e_r$, with minor impact on the recovered flow. 
Otherwise, the qualitative conclusions of BGA17 remain unaltered. 

\subsection{Posterior diagnostics}
\label{sec:Posterior_diagnostic}

We now define several diagnostics used to evaluate the quality and the consistency of our results.
We shall compare a quantity ${\bm x}$ (MF, SV, subgrid error, diffusion... in the spatial or spectral domain) with observations ${\bm x}^o$ (when available), or with the same quantities ${\bm x}^c$ from the CHAOS-6 geomagnetic model \citep{finlay2016recent}.
We define its time average 
\begin{linenomath*}\begin{eqnarray}
\label{eq:average_time_VO}
\hat{\bm x} = \frac{1}{t_f-t_i}\int_{t_i}^{t_f} {\bm x}(t)dt\,,
\end{eqnarray}\end{linenomath*}
with $t_i$ and $t_f$ the initial and final epochs, its ensemble mean 
\begin{linenomath*}\begin{eqnarray}
\label{eq:average_ensemble_VO}
\left< {\bm x}(t) \right> = \dfrac{1}{N_m}\sum_{k=1}^{N_m} {\bm x}^k(t)\,,
\end{eqnarray}\end{linenomath*}
the dispersion within the ensemble
\begin{linenomath*}\begin{eqnarray}
\label{eq:sigma_ensemble_VO}
{\bm \sigma}_x(t) = \displaystyle\sqrt{
\dfrac{1}{N_m-1}\sum_{k=1}^{N_m} \left( {\bm x}^k(t) - \left< {\bm x}(t) \right> \right)^2}\,,
\end{eqnarray}\end{linenomath*}
and finally the bias between our ensemble mean model and the reference ${\bm x}^c$, 
\begin{linenomath*}\begin{eqnarray}
\label{eq:delta_x_VO}
{\bm \delta}_{x}(t)=\displaystyle {\bm x}^c - \left< {\bm x}(t) \right> \,.
\end{eqnarray}\end{linenomath*}

We also define spatial power spectra of any magnetic trajectory ${\bf b}(t)$ as
\begin{linenomath*}\begin{eqnarray}
\label{eq:spectra_Mag_VO}
{\cal R}_b(n,t)=(n+1)\left(\frac{a^{\oplus}}{c}\right)^{2n+4}\sum_{m=0}^n \left[{g_n^m(t)}^2 +{h_n^m(t)}^2 \right]\,,
\end{eqnarray}\end{linenomath*}
with similar notations for $\dot{\bf b}(t)$, ${\bf d}(t)$ and ${\bf e}(t)$. 
$c=3485$ km is the Earth's core radius, and $g_n^m$ and $h_n^m$ are Schmidt semi-normalised spherical harmonic coefficients for the magnetic field at the Earth's surface.
Finally, the spatial power spectrum for core flow trajectories ${\bf u}$ writes
\begin{linenomath*}\begin{eqnarray}
\label{eq:spectrum_u_VO}
{\cal S}(n,t)=\frac{n(n+1)}{2n+1}\sum_{m=0}^n \left[
{{t_{c}}_n^m(t)}^2 + {{t_{s}}_n^m(t)}^2  + 
{{s_{c}}_n^m(t)}^2 + {{s_{s}}_n^m(t)}^2\right]\,,
\end{eqnarray}\end{linenomath*}
with ${{t_{c,s}}_n^m}$ and ${{s_{c,s}}_n^m}$ Schmidt semi-normalised spherical harmonic coefficients for the toroidal and poloidal components of the flow.
We also define the flow norm 
\begin{linenomath*}\begin{eqnarray}
\label{eq:norm_u}
{\cal N}=\sum_{n=1}^{n_u}\frac{n(n+1)}{2n+1}\sum_{m=0}^n \left[
{{t_{c}}_n^m}^2 + {{t_{s}}_n^m}^2  + 
{{s_{c}}_n^m}^2 + {{s_{s}}_n^m}^2\right]\,.
\end{eqnarray}\end{linenomath*}

The above power spectra can be considered for the ensemble mean or the dispersion within the ensemble, in which case they are respectively noted ${\cal R}_{<x>}(n,t)$ and ${\cal R}_{\delta x}(n,t)$.
Additionally, all those quantities may be averaged in time and/or computed only at analysis periods.
For example, the time-averaged spatial power spectrum of the dispersion of magnetic field solutions at analysis epochs is $\hat{\cal R}_{\delta b}^a(n)$.
The same convention as above holds for core flow spectra.

\section{Results}
\label{sec:Results}

We apply our algorithm to VO and GO magnetic field observations over a period spanning from $t_i=1996.92$ to $t_f=2016.92$.
We recall that since we use satellite measurements from CHAMP and \textit{Swarm} missions, VOs are available only over the period 2000--2010 and 2014--2017, whereas GOs are available over the whole time-span.
Analysis are performed every $\Delta t^a=4$ months. 
The sequences of analyses and forecasts between 1997 and 2001 are used to warm-up the filter (see Figure 7 in BGA17), avoiding an increase in the ensemble spread over the first years of the targeted satellite era.  
This warm-up period is not considered below when interpreting the ensemble of inverted magnetic field and flow. 
We first describe predictions from our re-analysis for observations in the physical domain (\S\ref{sec:Physical_space}), before we present the resulting magnetic model (\S\ref{sec:Geomag_model}), and insights on core flows over various time-scales (\S\ref{sec:Core_flows}). 

\subsection{Geomagnetic field models}
\label{sec: mag}

\subsubsection{Predictions for GO and VO series}
\label{sec:Physical_space}

\begin{figure*}
\centering{
	\includegraphics[width=1\linewidth]{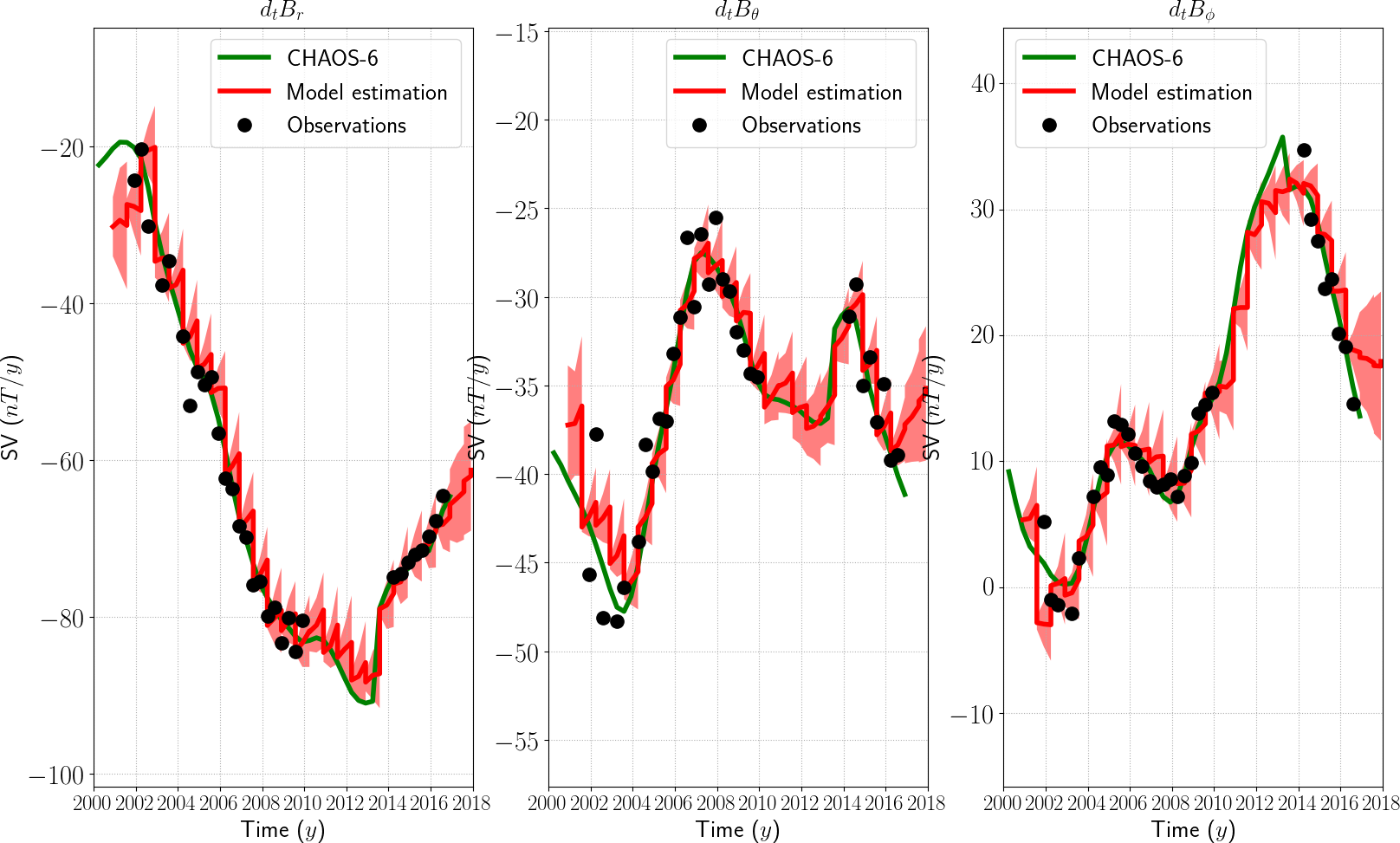}
	\includegraphics[width=1\linewidth]{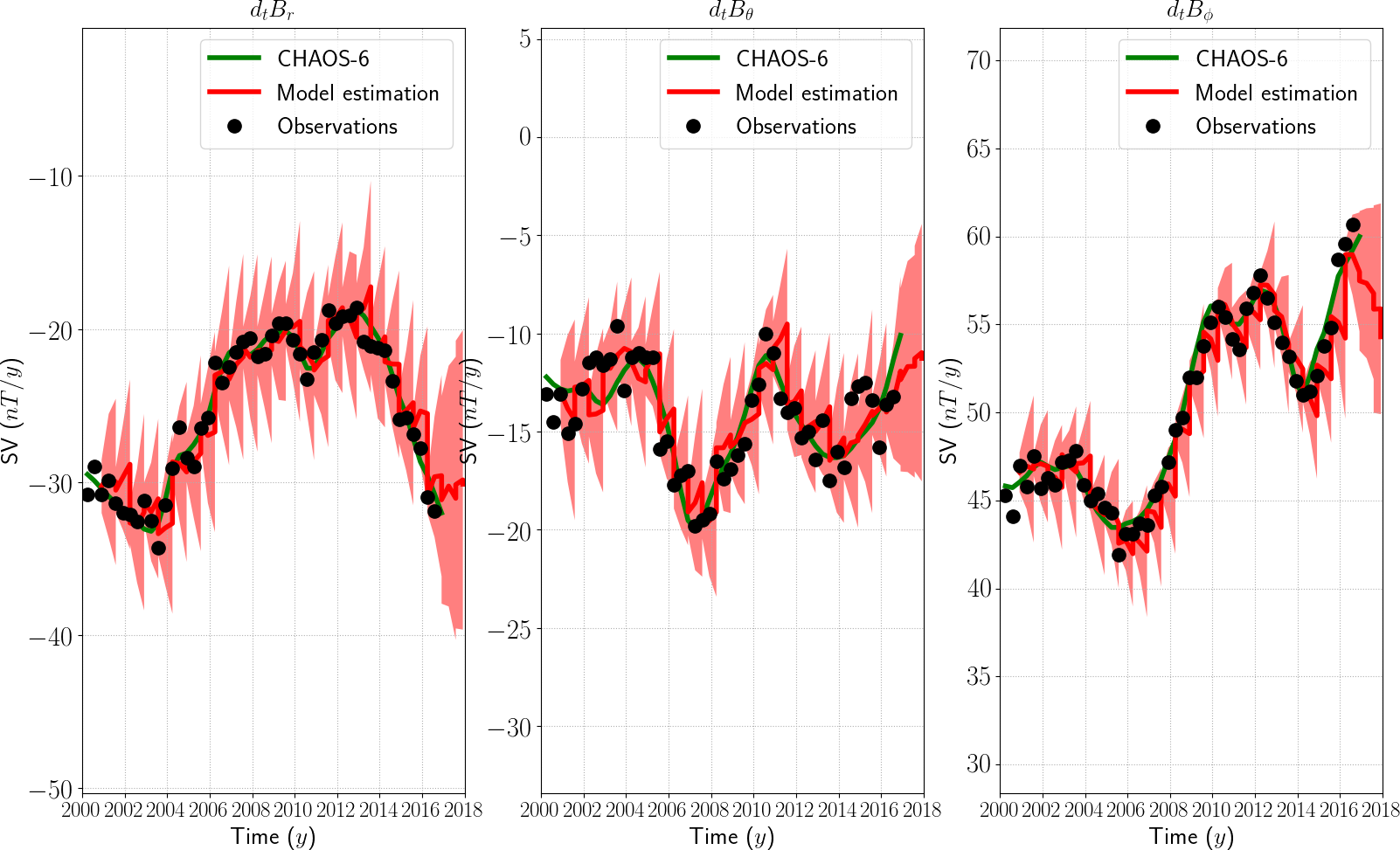}}
	\caption{
	SV time series for the three components $(dB_r/dt, dB_\theta/dt, dB_\phi/dt)$, at one VO location $\left\lbrace r=6671\; km, \theta=90^{\circ}, \phi=88,8^{\circ} \right\rbrace$ (top), and at Chambon-la-for\^et $\left\lbrace r=6366 \; km, \theta=42^{\circ}, \phi=2^{\circ} \right\rbrace$ (bottom). 
SV observations are shown in black, CHAOS-6 predictions in green, predictions from our analysis in red. 
The shaded area correspond to $\pm \sigma_{\dot{b}}$, see equation (\ref{eq:sigma_ensemble_VO}).
	}
	\label{fig:R15_VO_90_88+GO_CLF}
\end{figure*}

We compare in Figure~\ref{fig:R15_VO_90_88+GO_CLF} our series of SV forecasts and analysis with two examples of observation series (one VO and one GO), and with the predictions from CHAOS-6.
The large spread of the SV forecasts is to be expected given the large uncertainties associated with subgrid errors and the large-scale flow (see BGA17). 
At both sites, the dispersion within the ensemble of SV trajectories encompasses most of the time the observations.
Moreover the predictions from CHAOS-6 and from our ensemble of SV models are generally consistent.
Our algorithm thus seems able to provide a coherent estimate of the SV probability density function (PDF) at the Earth's surface and at satellite altitude.
In addition, we highlight that even during the period 2010-2014 where no VO data are available, the trajectory of SV model, controlled by the stochastic prior and GO data only, remains reasonable, with a slight increase in the ensemble spread that always contains CHAOS-6.
Note that our algorithm tends to drive the system toward low SV values (see the saw-tooth patterns in Figure \ref{fig:R15_VO_90_88+GO_CLF}).
This feature is to be expected given our choice of the stochastic models for ${\bm u}_H$ and $e_r$, which control the evolution of the SV. 
In the absence of data constraints, the process will drift back the ensemble average trajectories for ${\bm u}_H$ and $e_r$ towards the average dynamo state, which by construction is responsible for a weak SV.
This is not a major drawback as soon as we analyse frequently enough, though it does limit the prediction capabilities of our tool (as discussed in BGA17).

\begin{figure*}
\centerline{
	\includegraphics[width=0.33\linewidth]{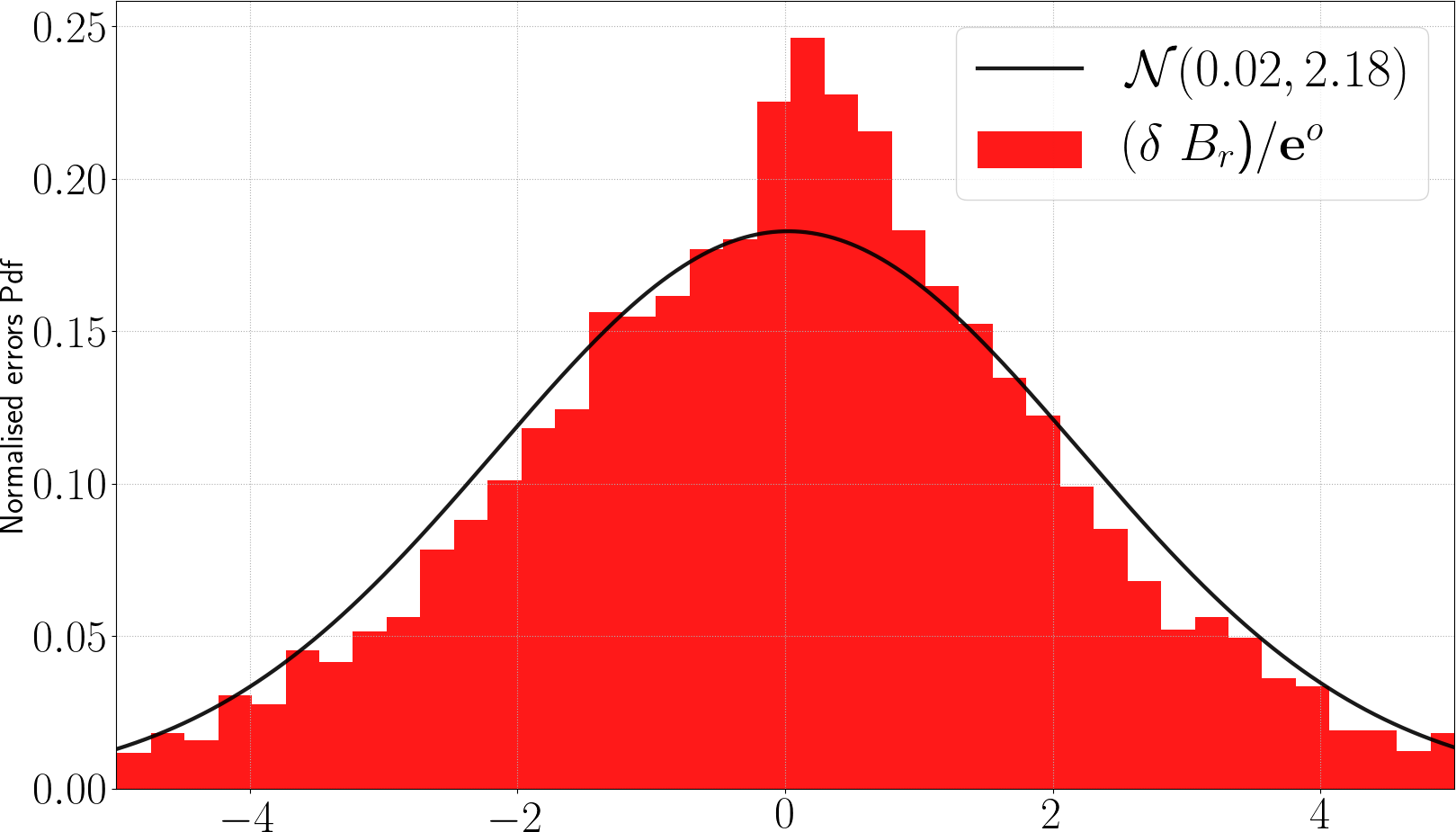}
	\includegraphics[width=0.33\linewidth]{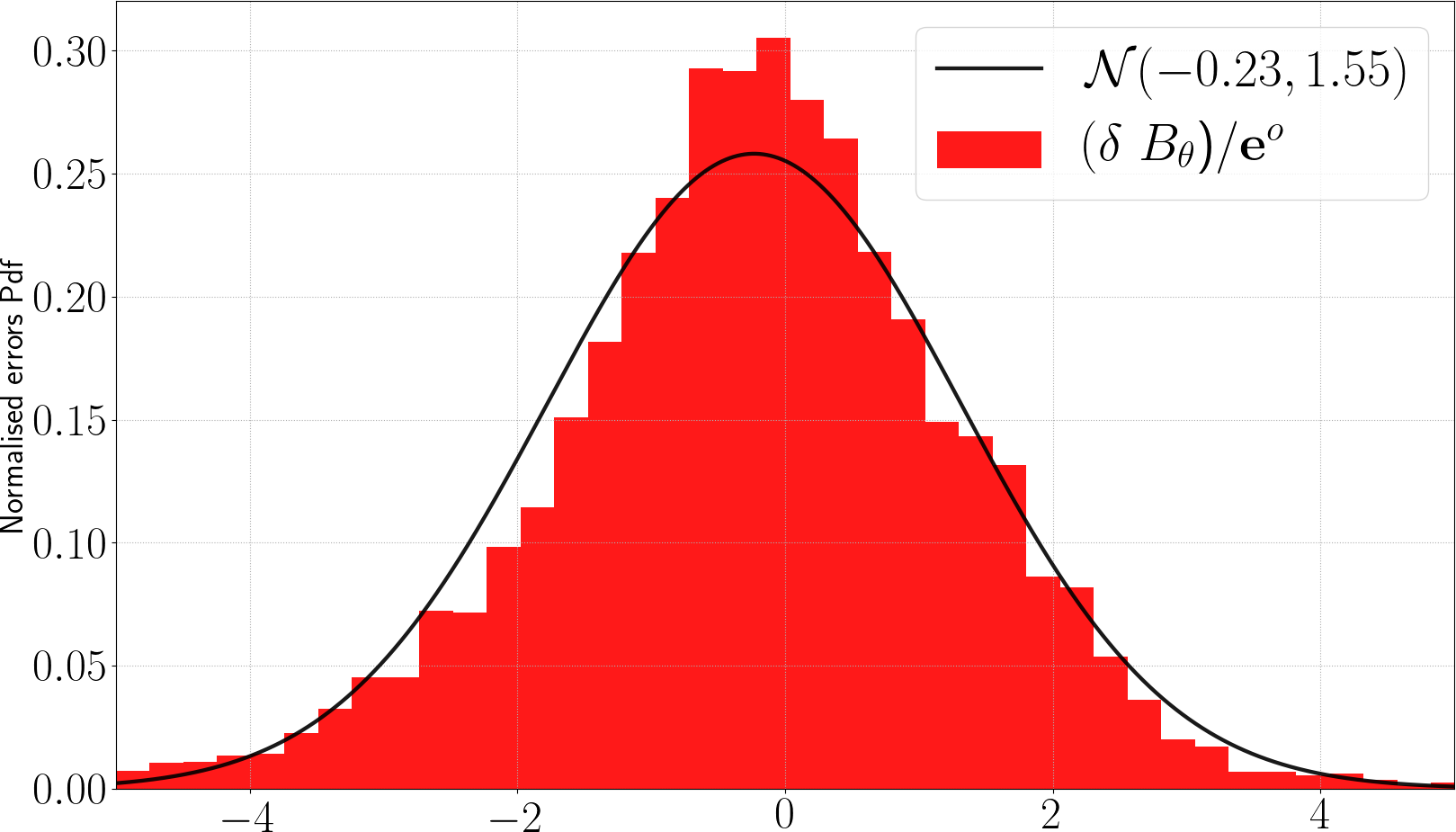}
	\includegraphics[width=0.33\linewidth]{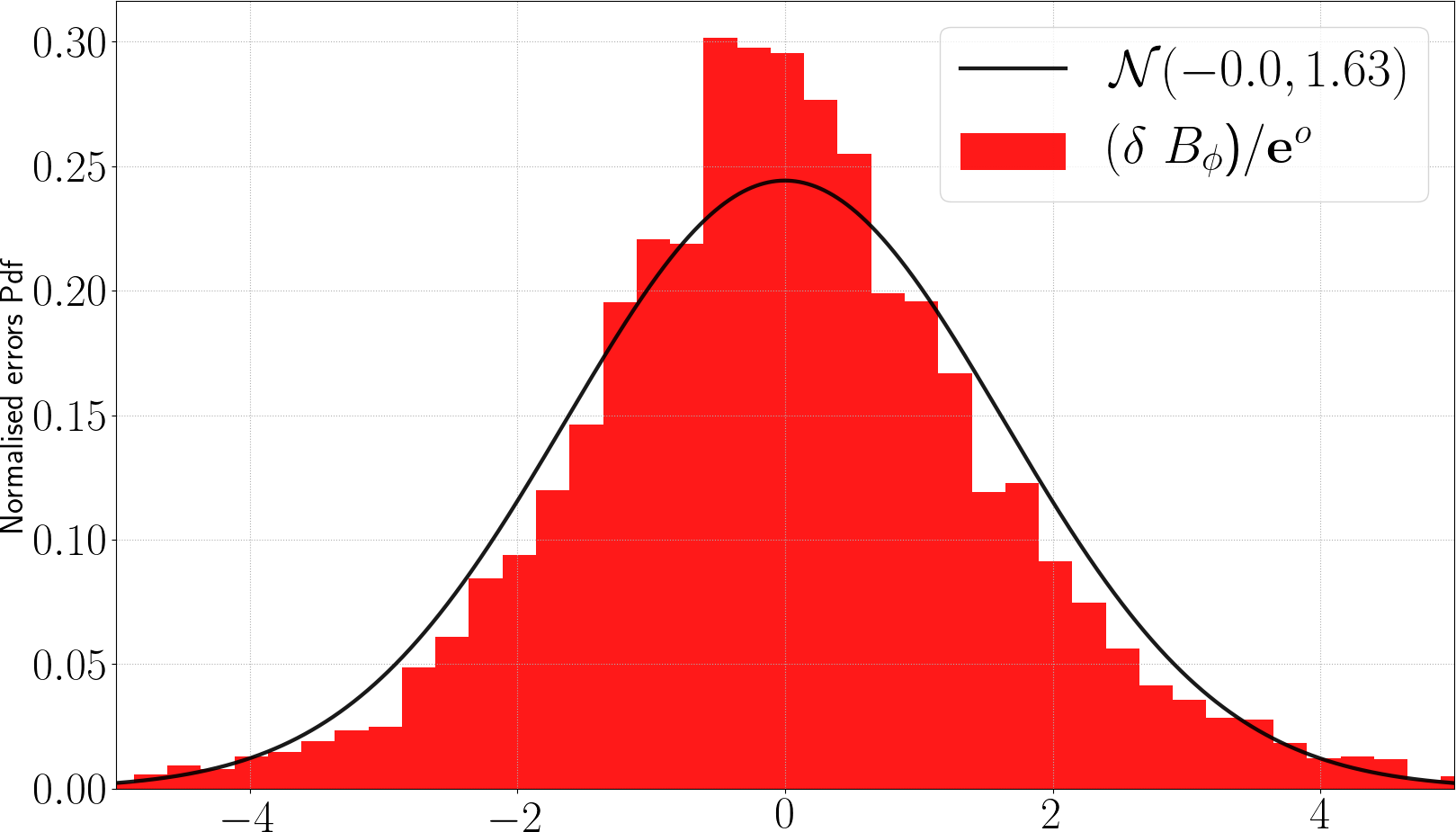}}
\centerline{
	\includegraphics[width=0.33\linewidth]{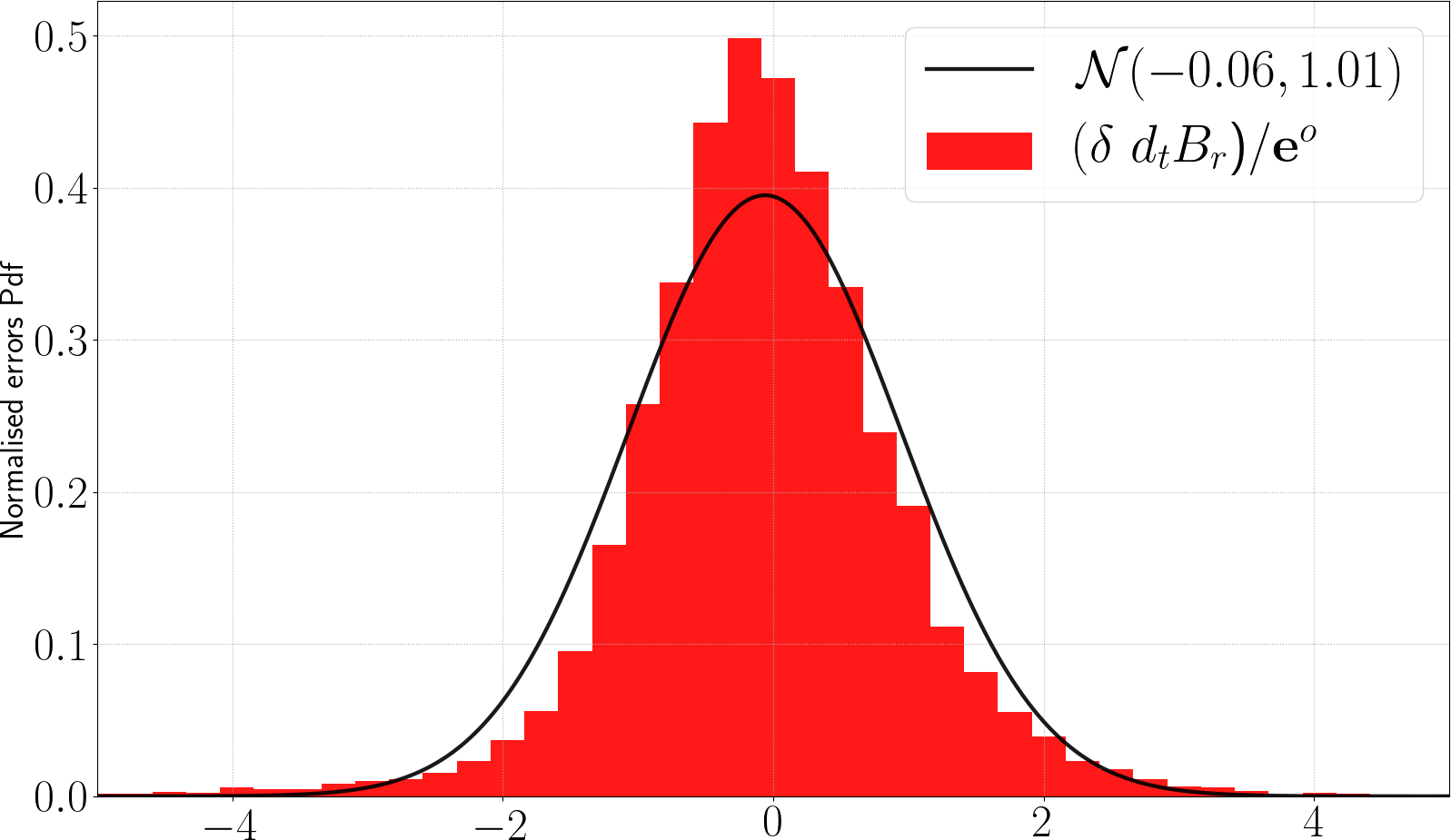}
	\includegraphics[width=0.33\linewidth]{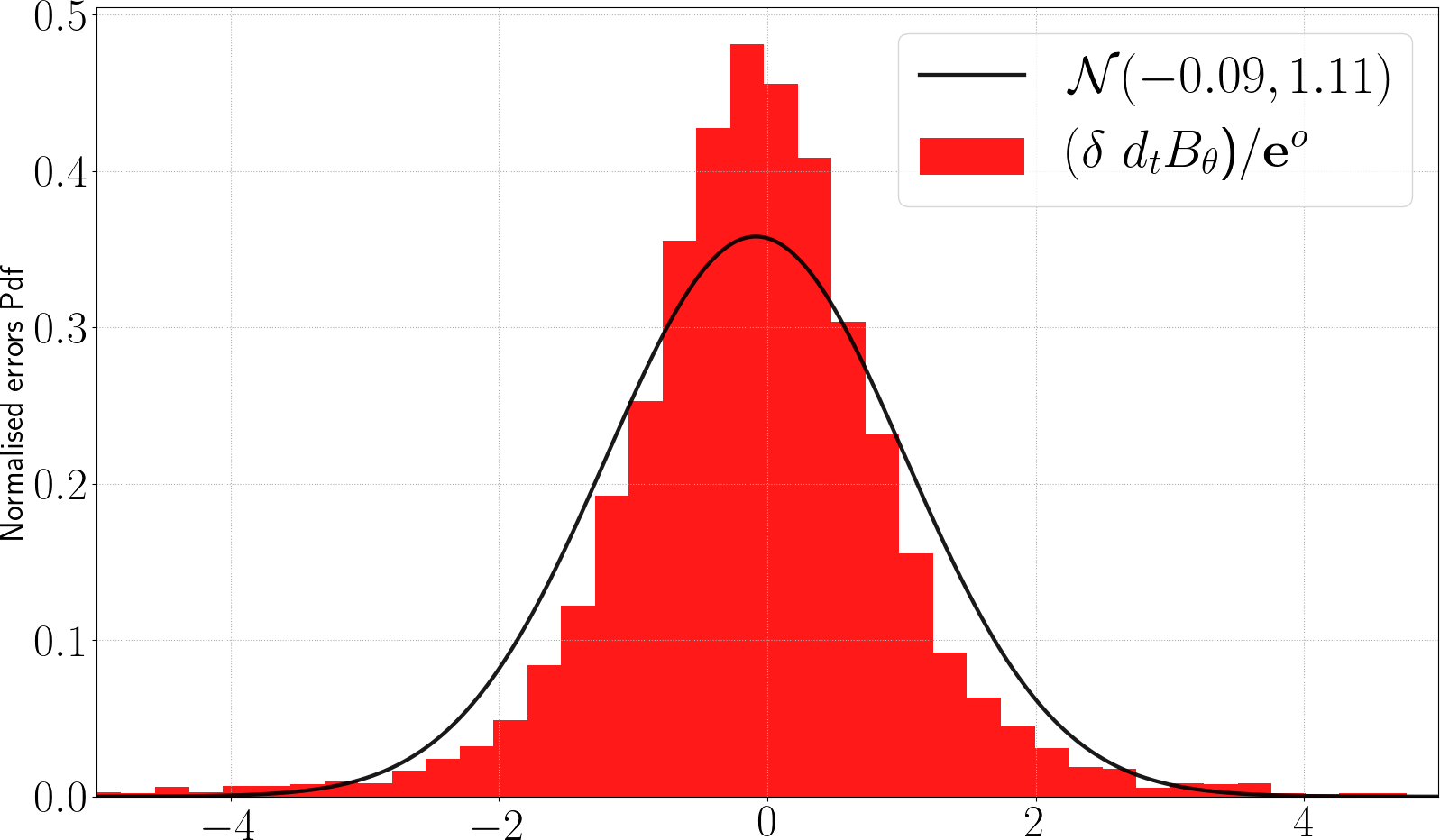}
	\includegraphics[width=0.33\linewidth]{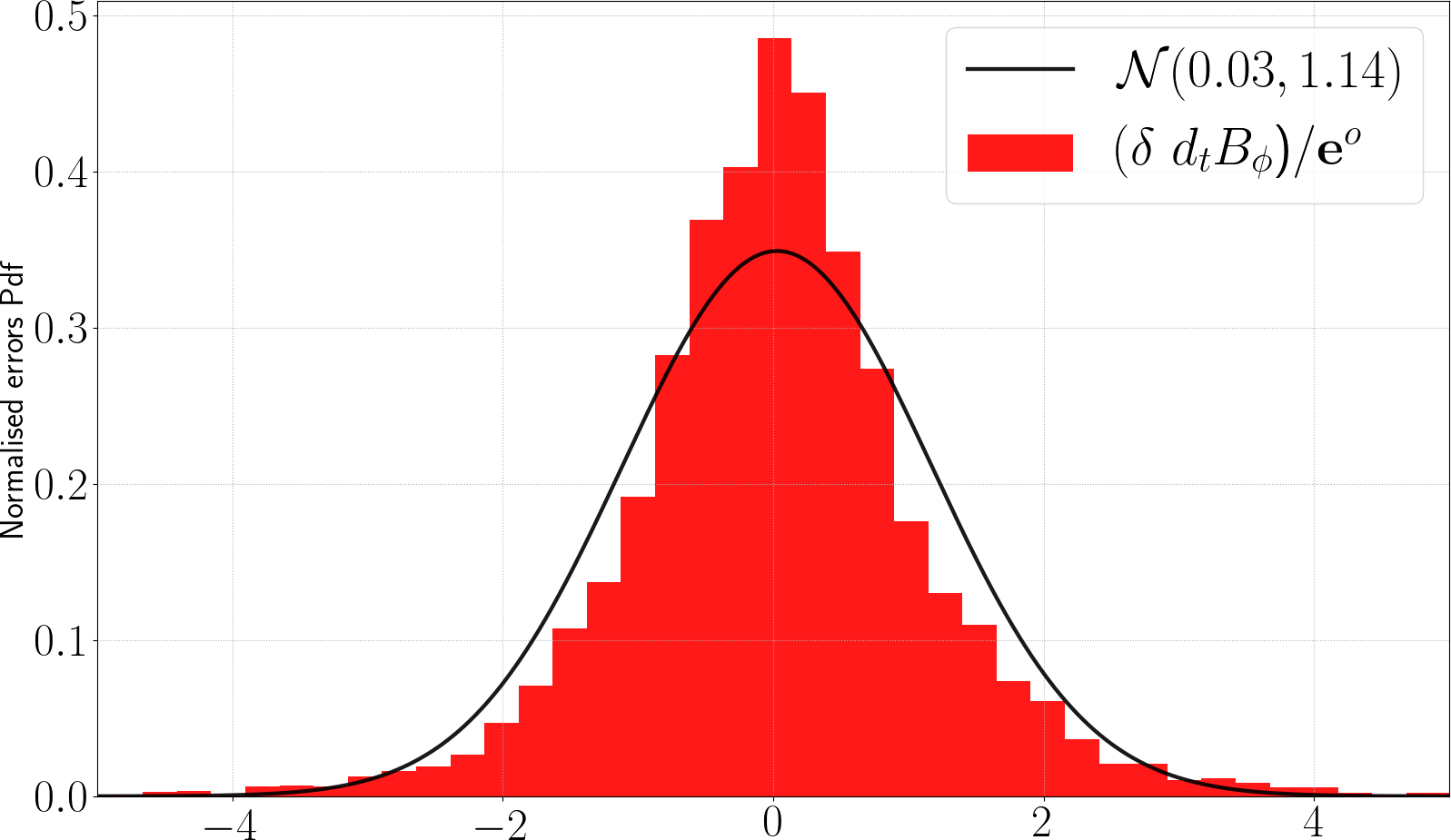}}
	\caption{
	Top: Histograms of MF prediction errors $\delta_{MF}$ (eq. \ref{eq:delta_x_VO}), accumulated over all analysis epochs, normalised to the observation errors, for the components $B_r$ (left), $B_\theta$ (middle) and $B_\phi$ (right). 
Superimposed in black are the Gaussian distribution fits obtained with the mean $\mu$ and the variance $\sigma^2$ for each of the three distributions.
	Bottom: same histograms for the SV prediction errors $\delta_{SV}$.
	}
	\label{fig:R15_hist_Data-Model}
\end{figure*}

We check in Figure~\ref{fig:R15_hist_Data-Model} the accuracy with which our model fits MF and SV observations, with the histograms of the prediction errors (over all analyses) normalised to the observation errors, for the three components of the magnetic field.
Concerning the MF, prediction errors are only weakly biased, excepted for $B_{\theta}$ (normalised biases on the three components are $\mu_r = -0.02$, $\mu_\theta = -0.23$ and $\mu_\phi = 0.0$).
The histograms of prediction errors are reasonably close to Gaussian for the three components with observation errors that appear to be under-estimated on average, in particular on $B_r$ (normalised r.m.s. errors on the three components are $\sigma_r = 2.18$, $\sigma_\theta = 1.55$ and $\sigma_\phi = 1.63$).
The SV predictions errors are remarkably consistent with the a priori errors with small biases and standard deviation close to unity for the three components ($\mu_r = -0.06$, $\sigma_r = 1.01$; $\mu_\theta = -0.09$, $\sigma_\theta = 1.11$ and $\mu_\phi = 0.03$, $\sigma_\phi = 1.14$), even though the distributions appears more peaked than a Gaussian.
The Kalman filter employed here implicitly assumes Gaussian distributed data errors. 
However, the above remark suggests that alternative treatments of data residuals may be worth considering in future studies \citep[e.g. L1 or Huber norms, see][]{constable1988parameter,farquharson1998non}.

\subsubsection{Field models, and contributions to the SV}
\label{sec:Geomag_model}

\begin{figure*}
\centerline{
	\includegraphics[width=0.5\linewidth]{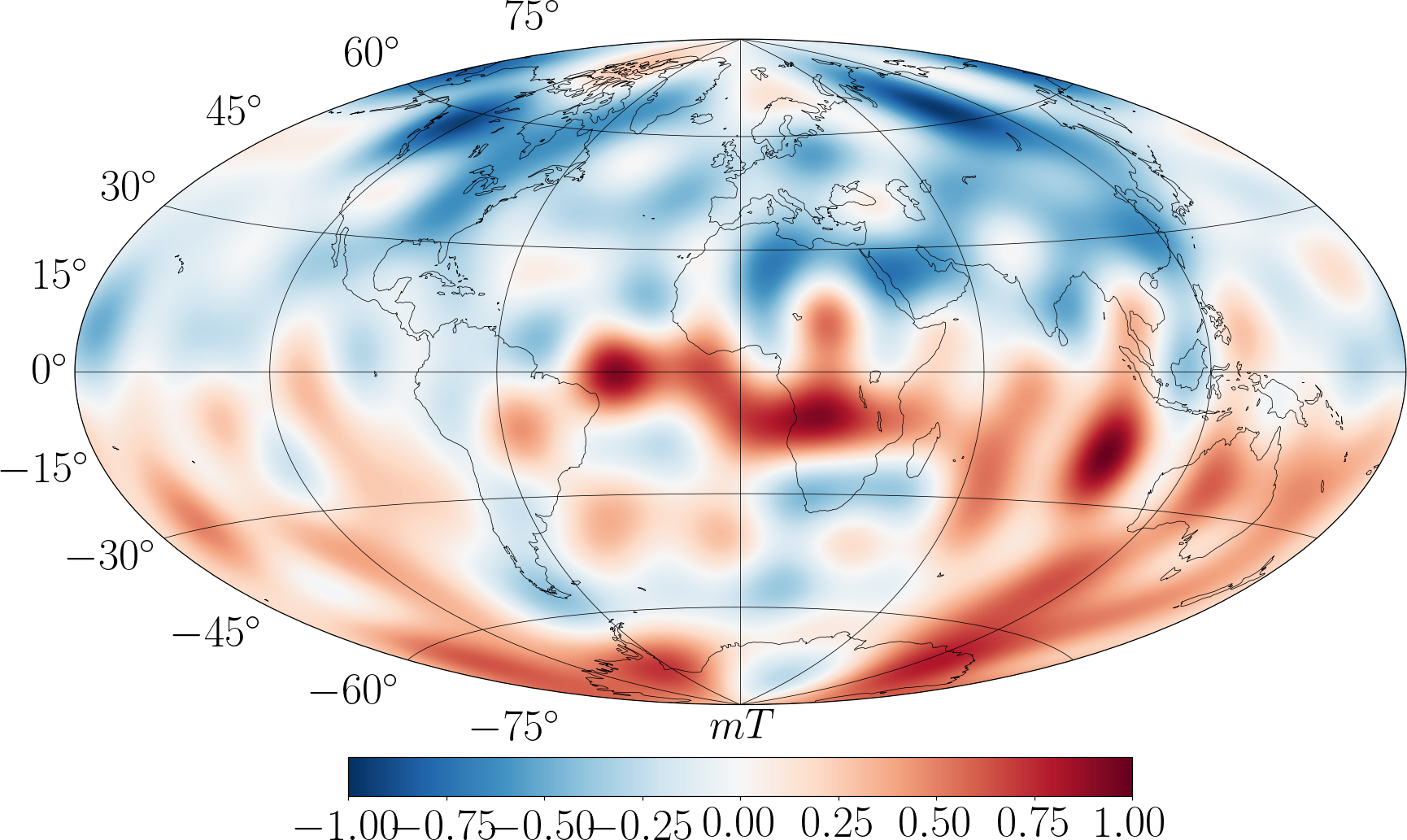}
	\includegraphics[width=0.5\linewidth]{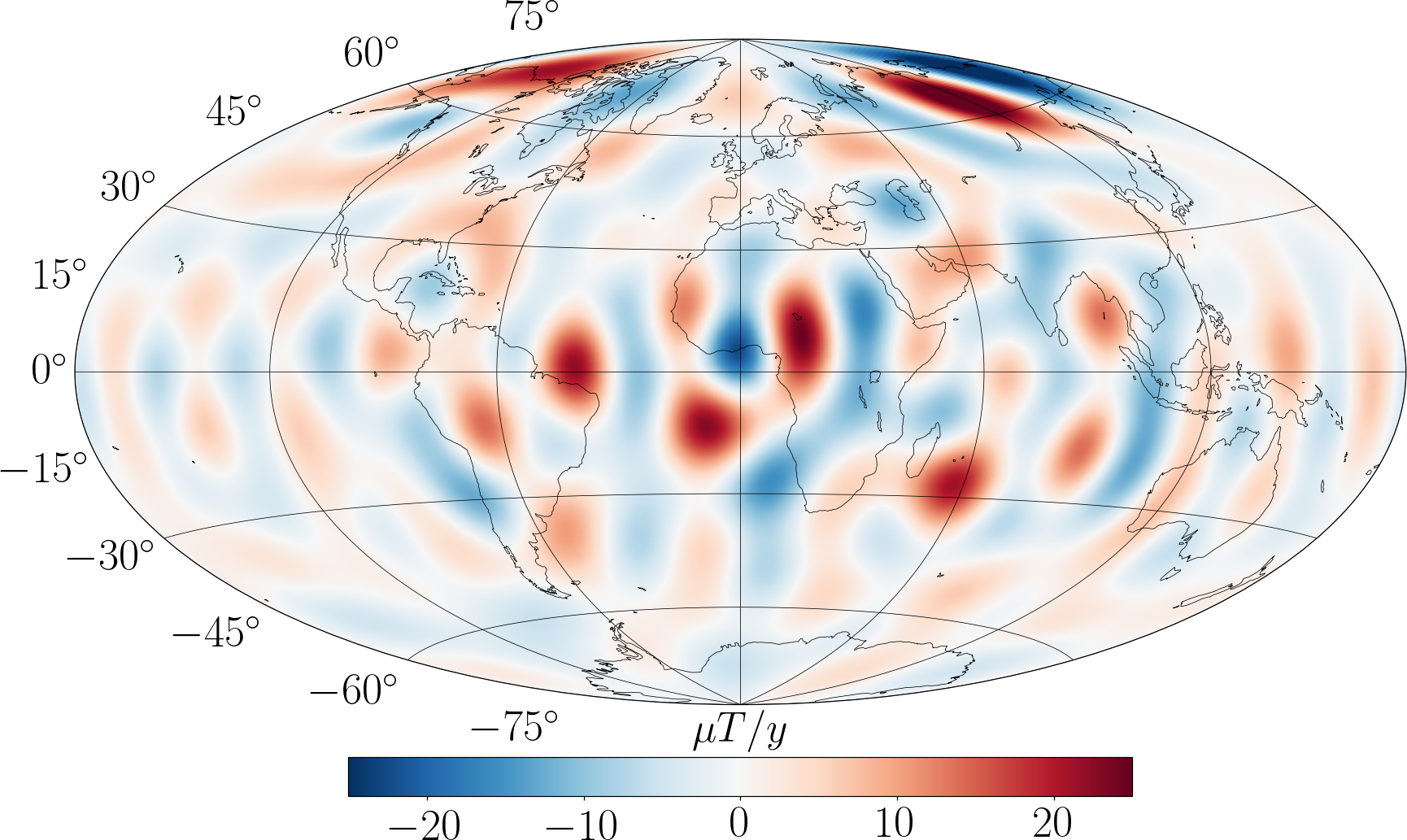}}
\centerline{		
	\includegraphics[width=0.5\linewidth]{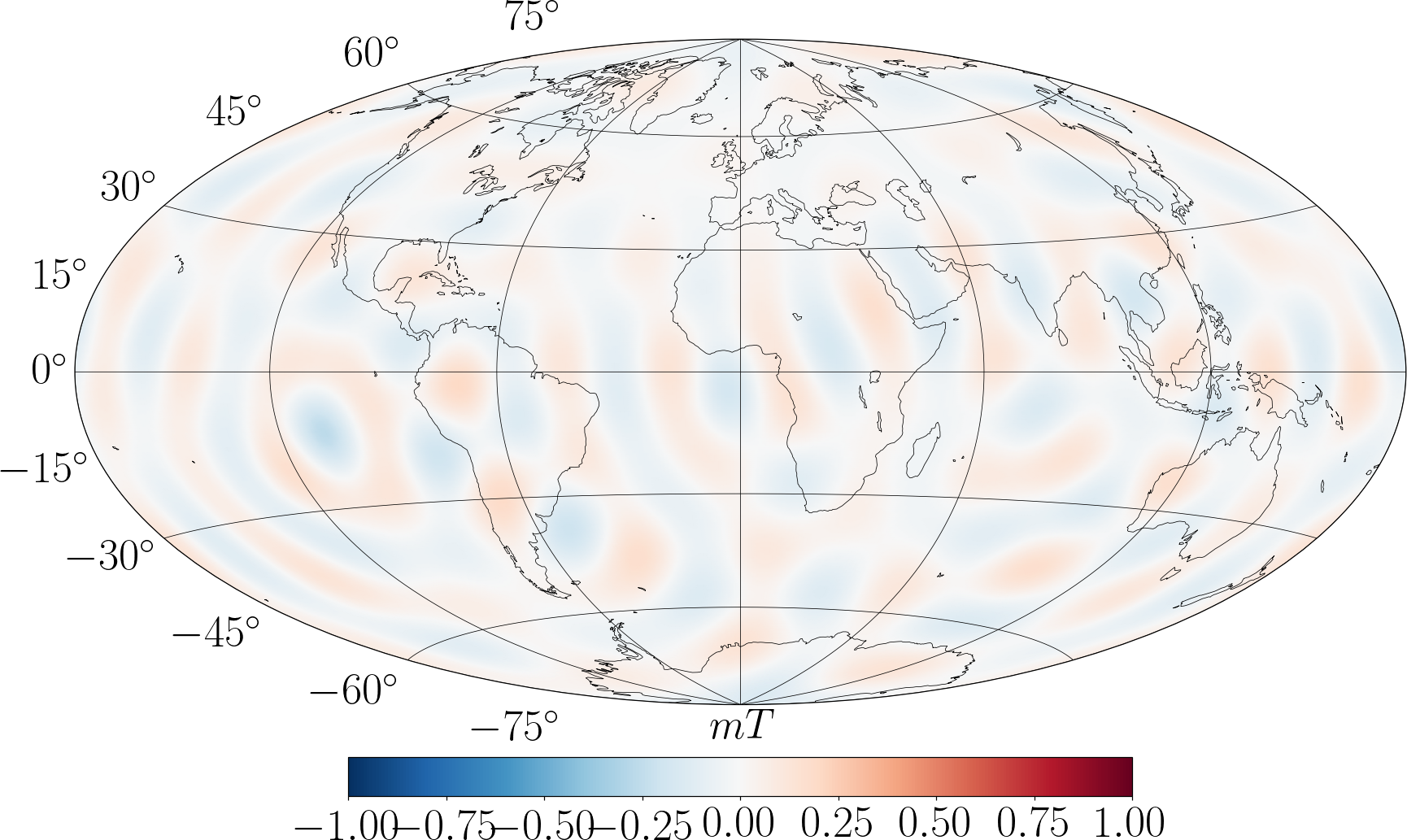}
	\includegraphics[width=0.5\linewidth]{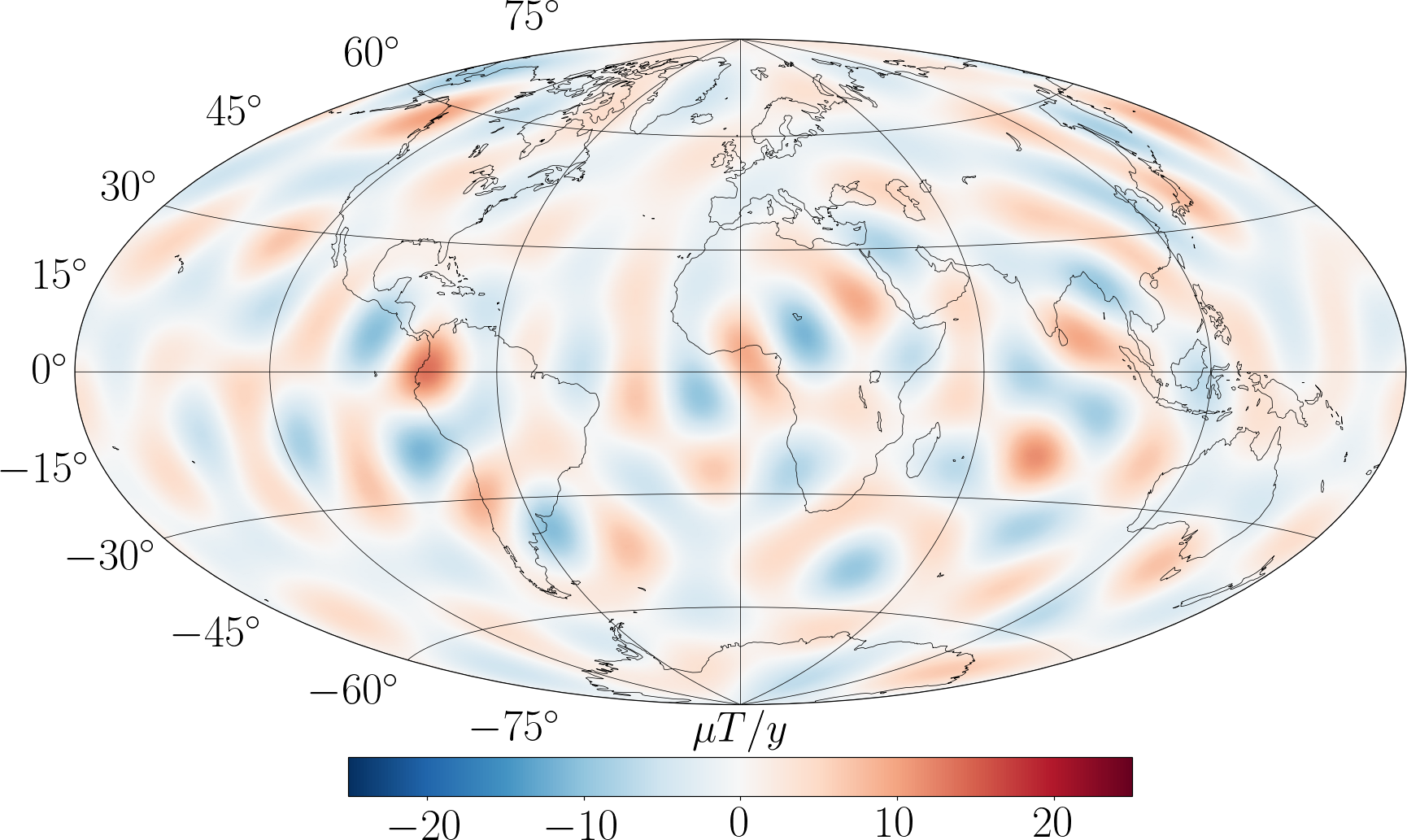}}
	\caption{
	Top: CMB maps of the ensemble average radial magnetic field $\left< {\bf b} \right>$ (eq. \ref{eq:average_ensemble_VO}) in 2017 (left: MF in mT; right: SV in $\mu$T/yr), as estimated with our algorithm. 
Bottom: MF (left) and SV (right) maps of the difference of our ensemble average field with CHAOS-6 (truncated at degree 14) at the CMB (with the same colorscales).
	}
	\label{fig:R15_geomag_Mod_CHAOS6}
\end{figure*}

We now describe in more detail our MF and SV models. 
We present in Figure~\ref{fig:R15_geomag_Mod_CHAOS6} MF and SV maps for our ensemble average model at the CMB truncated at spherical harmonic degree $n=14$.
Comparing it to a more traditional field model CHAOS-6, which is temporally regularised, the overall agreement is very good, indicating that our tool is indeed capable of producing reasonable field models. 
MF discrepancies to CHAOS-6 are relatively small, with peak to peak values less than $10\%$ of the total amplitude for a field truncated at degree 14.
They are dominated by isotropically distributed, small length-scale patterns. 
As well as being dominated by small length-scales, the disagreements are larger for the SV, with peak to peak differences about $30\%$ of the total amplitude, which is to be expected given the blue SV spectrum at the CMB, meaning that small length scales dominate.
Interestingly, the largest differences are localised under South America and the Indian Ocean, where the planetary gyre respectively detaches from and joins the equatorial belt \citep{pais2008quasi} and where rapid time-dependence is observed \citep{finlay2016gyre}.

\begin{figure*}
\centerline{
	\includegraphics[width=0.5\linewidth]{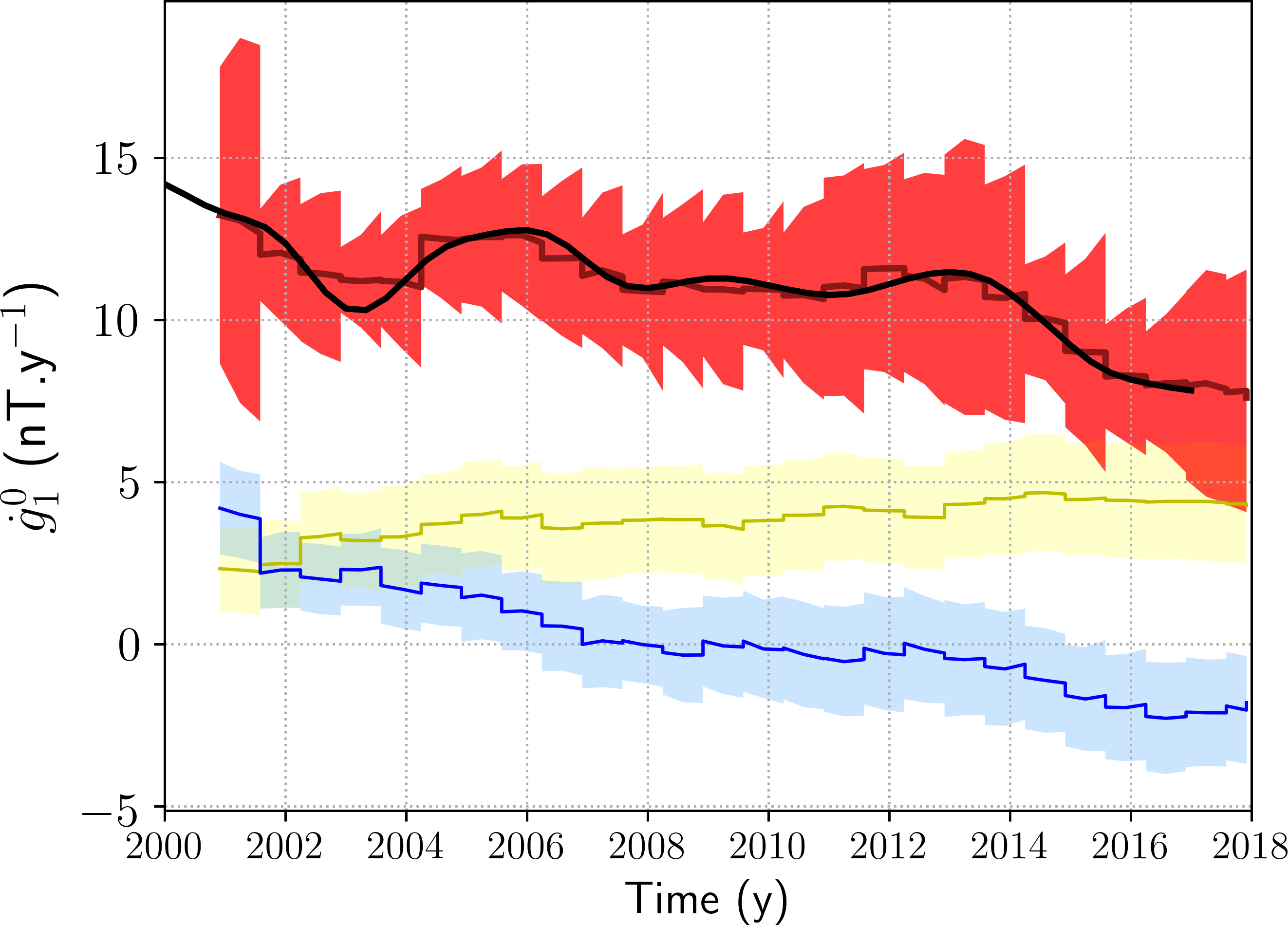}
	\includegraphics[width=0.5\linewidth]{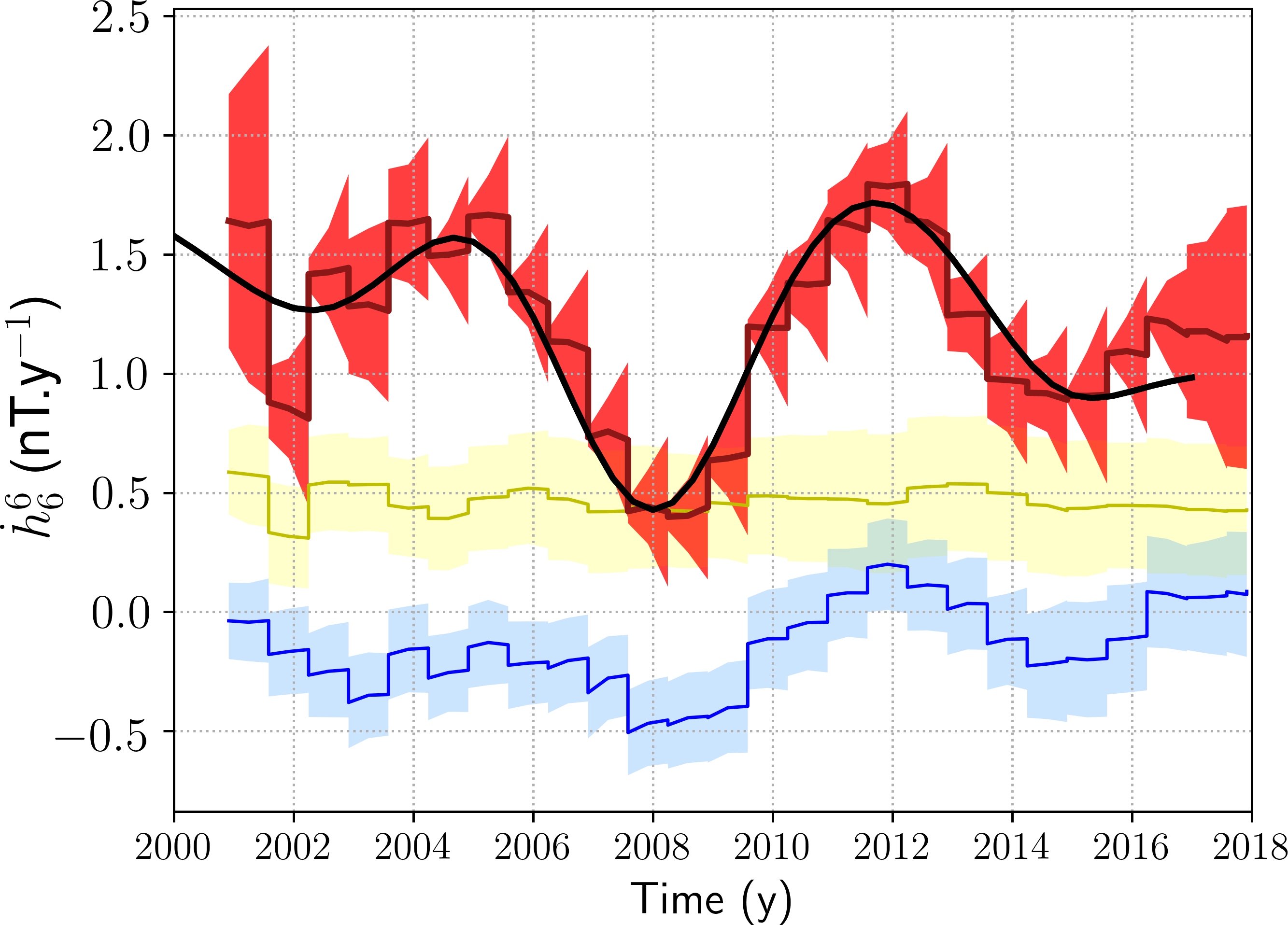}}
	\caption{
	SV spherical harmonic coefficient time series for $\dot{g}_{1}^{0}$ (left) and $\dot{h}_{6}^{6}$ (right). Predictions from our ensemble average model are shown in dark red ($\pm 2 \sigma_{\dot{b}}$ in red), and CHAOS-6 in black. Contributions from subgrid errors and diffusion extracted from our ensemble of realizations are superimposed in respectively blue and yellow (with dispersions $\pm 1 \sigma_{diff}$ and $\pm 1 \sigma_{er}$ in the corresponding colors).
	}
	\label{fig:R15_dg1-48}
\end{figure*}

In Figure~\ref{fig:R15_dg1-48} we show the various contributions in our model to two SV spherical harmonic coefficient series. 
The dispersion within the ensemble of models is large enough to include time changes as estimated by CHAOS-6, with some exceptions during the high solar activity era, e.g. in 2002 for $h_6^6$, and at the very end of the CHAOS-6 era (this latter possibly in link with the damping of SA towards end-points in the regularised field model).    
We notice a larger spread of the analysis for the axial dipole than for non-zonal coefficients of intermediate length-scale such as $h_6^6$.
This may be a consequence of the weaker constraint on zonal coefficients from surface observations \citep[e.g.][]{kotsiaros2012}, although we only notice such behaviour for $g_1^0$.
An enhancement of the dispersion is noticeable between 2010 and 2014, displaying in the spectral domain the impact of the decreasing number of data during this era when no vector satellite data were available.
Over 2001--2006, the ensemble average $h_6^6$ trajectory shows distinctive square shaped variations, probably partly related to variations in the number of data satisfying selection criteria during this interval of enhanced solar activity when only CHAMP data were available.

\begin{figure*}
\centering
{	\includegraphics[width=0.7\linewidth]{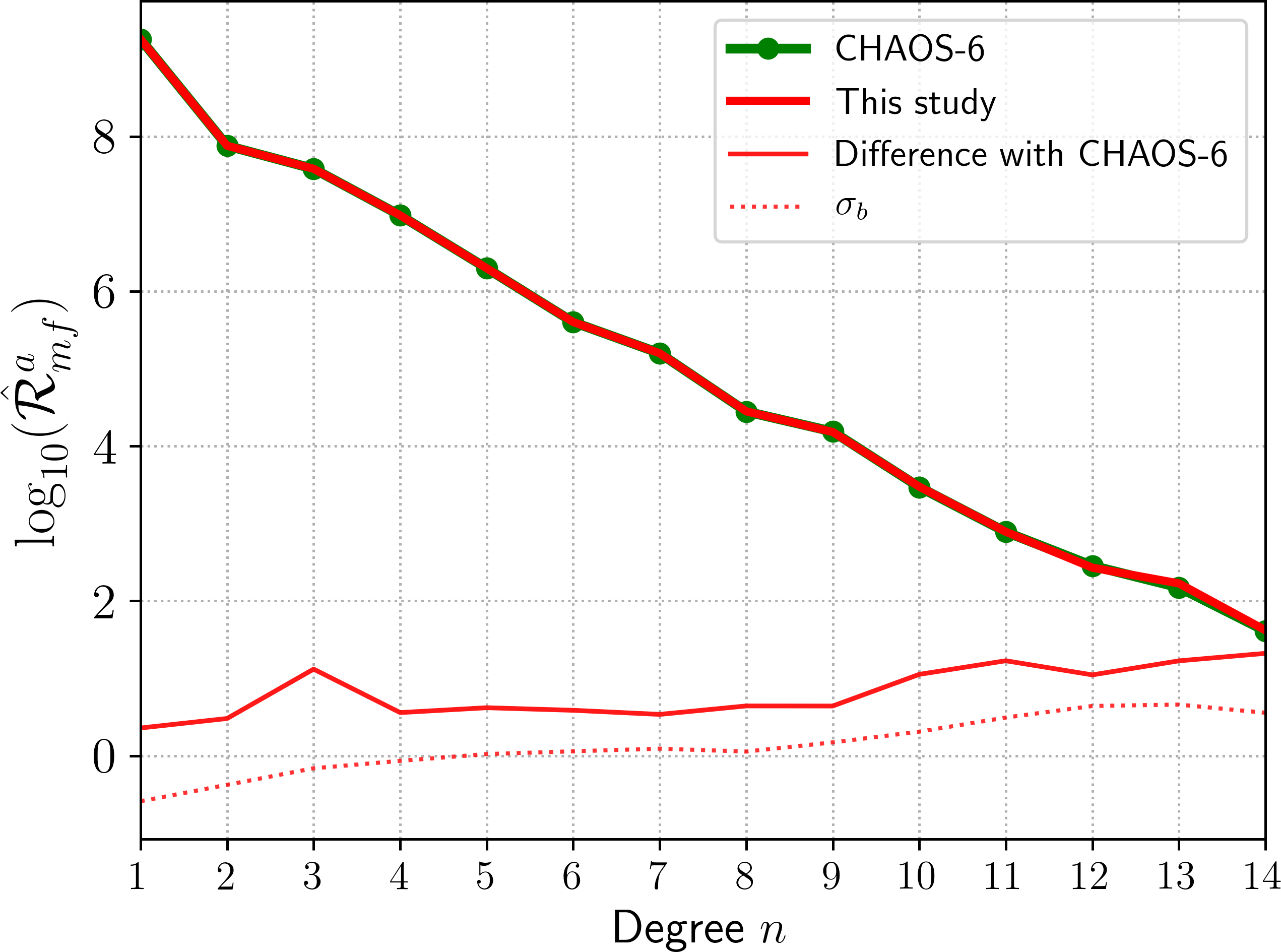}
	\includegraphics[width=0.7\linewidth]{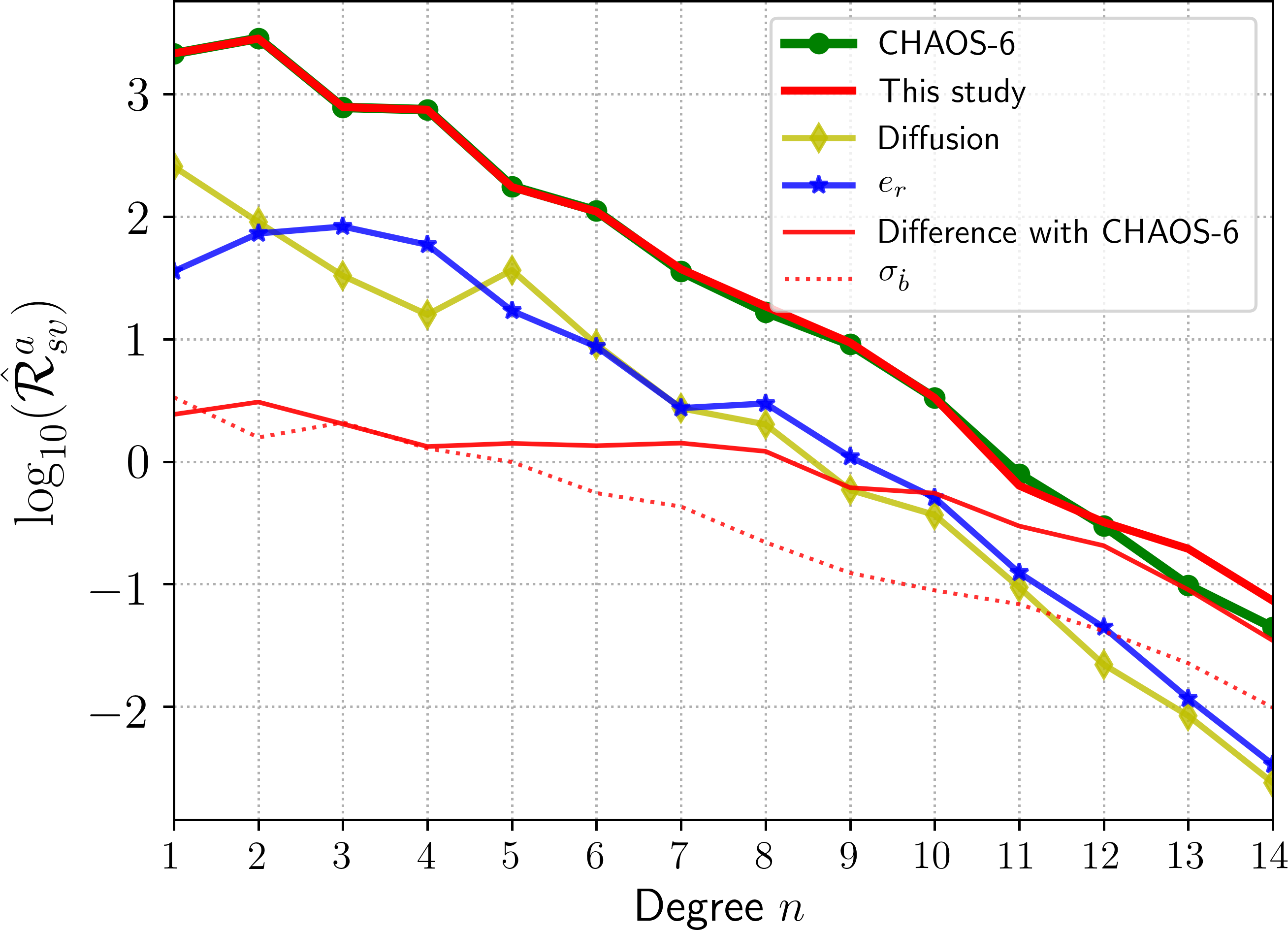}}
	\caption{
	Top: time averaged spatial power spectra at the Earth surface of the magnetic field of CHAOS-6 ($\hat{\cal R}_{b^c}^a$, eq. (\ref{eq:spectra_Mag_VO}), in green), our estimate ($\hat{\cal R}_{\left< b \right>}^a$, in red), the difference between the two ($\hat{\cal R}_{\delta b}^a$, red thin line) and the dispersion within our ensemble of analyses ($\hat{\cal R}_{\sigma_b}^a$, dotted line). 
Bottom: \textit{idem} for the SV, superimposed with the spectra of the contributions from subgrid errors (blue) and from diffusion (yellow). 
	}
	\label{fig:R15_spectre_CHAOS6-Model}
\end{figure*}

Spatial spectra shown in Figure~\ref{fig:R15_spectre_CHAOS6-Model} summarise the characteristics of our model in the spectral domain.
We find excellent agreement with CHAOS-6 for the main field and its secular variation, except at the small length scales of the SV ($n>10$), which are more likely to be affected by the different data set chosen and by the different temporal kernel used (short time windows in our case against whole time-span for CHAOS-6).
The ensemble spread gives a good approximation of the characteristic distance between our model and CHAOS-6.
Diffusion and subgrid errors in the SV have approximately the same amplitude except for the dipole. 
The power stored in these two SV sources represents about $10$ to $20\%$ of the total SV energy at all scales.

Even though the dispersion within the model predictions is large enough to encompass most of the MF and SV observations, the dispersion within the ensemble of realizations is lower, by a factor about 2.5, than the distance between the ensemble average model and CHAOS-6 for both the MF (at all length-scales) and the SV (towards small length-scales only). 
A complete account of SV errors from all subgrid interactions \citep[see][]{baerenzung2017modeling} may help reduce the above under-estimation.
Our current estimate is nevertheless larger than that obtained for the COV-OBS.x1 model \citet[see Figure 4 in][the error spectrum in 2010]{gillet2015stochastic}.
We suspect that the accumulation of data (assumed independent) during the construction of this latter field model involved too strong a decrease of the posterior error within the COV-OBS framework.
The more consistent approach to error propagation developed here and presented in Figure \ref{fig:R15_spectre_CHAOS6-Model} favours larger uncertainties on spherical harmonic coefficients during the satellite era.

\begin{figure*}
\centerline{
	\includegraphics[width=0.7\linewidth]{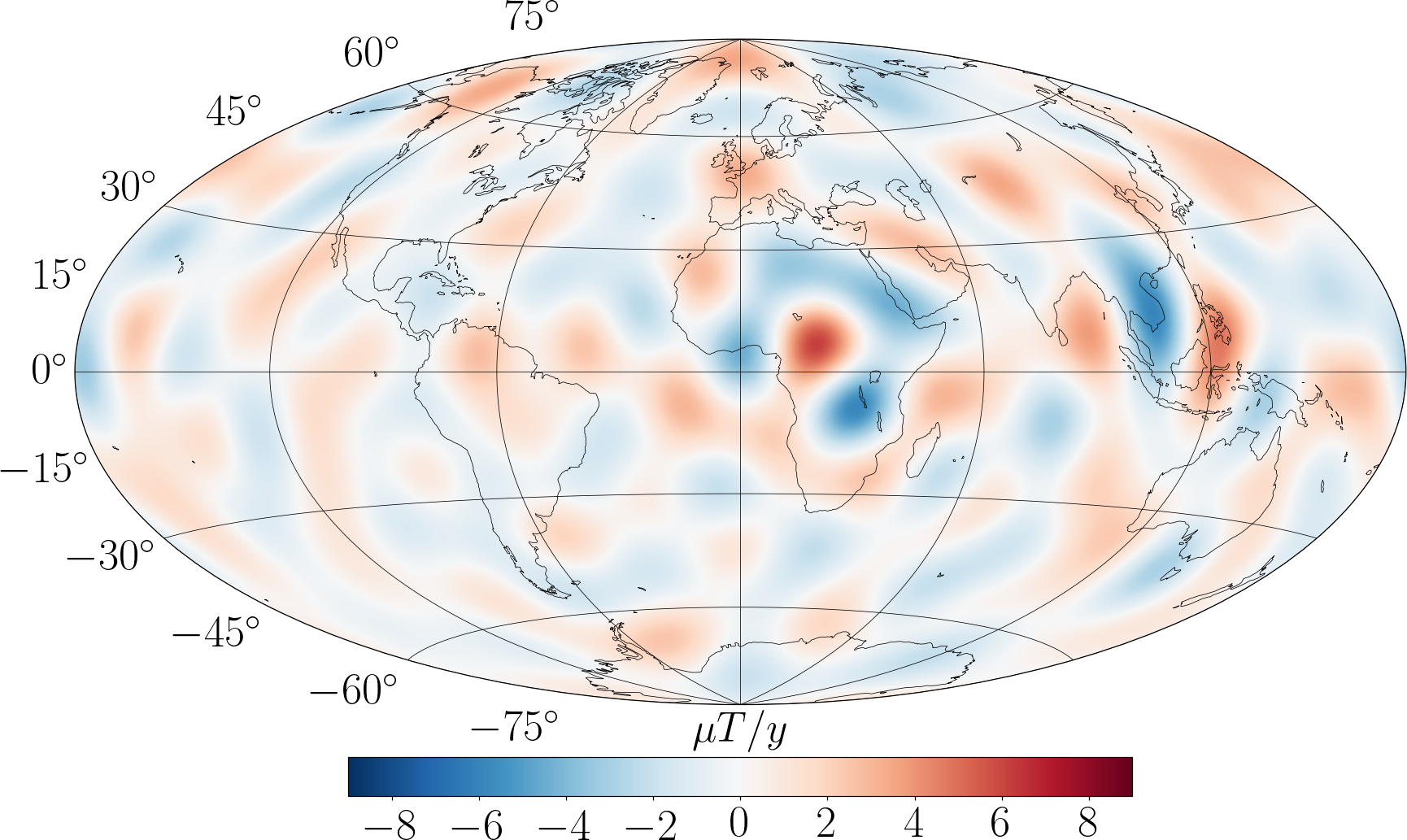}}
\centerline{
	\includegraphics[width=0.7\linewidth]{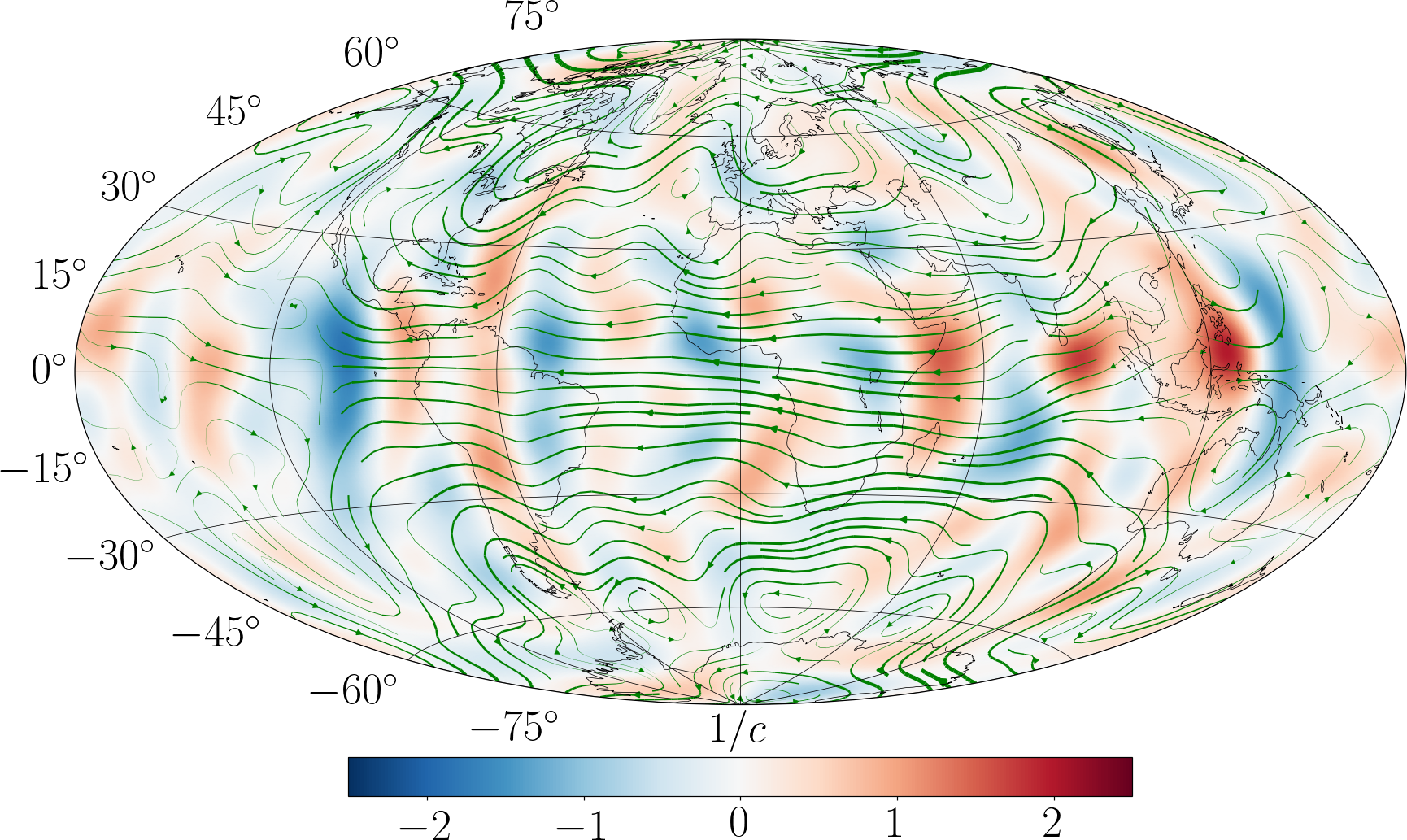}}
	\caption{
Magnetic diffusion at the CMB (top, color-scale in $\mu$T.y$^{-1}$), 
and horizontal divergence $\nabla_{h}\cdot{\bf u}_h$ (bottom, color-scale in $10^{-3}$ yrs$^{-1}$) superimposed with passive tracers trajectories (black, tracer size scale in km/yr), for the ensemble average model in 2017. 
Core flow visualisations are performed using the tools provided at {\tt https://geodyn.univ-grenoble-alpes.fr/}.
The size of the tracers is proportional to the velocity field (see the legend).
The initial positions of the tracers is random; each trajectory is advected by the velocity field for a fixed time; along each trajectory, the late (early) positions are darker (lighter). 
	}
	\label{fig:R15_Maps_Diff}
\end{figure*}

Overall, we are generally able to retrieve earlier well-established results. 
For instance the contribution from advection dominates (over diffusion) the axial dipole decay \citep{finlay2016recent,barrois2017contributions} and its fluctuations -- even though our estimate for the contribution from diffusion to $dg_1^0/dt$, shifted upward by a couple of nT/yr in comparison with the results of BGA17 (see \S\ref{sec: assim}), amounts to a relatively larger fraction over the latest years where the dipole decay tends to be weaker.
The ensemble average SV originating from diffusion is presented in Figure~\ref{fig:R15_Maps_Diff} for 2017: the most significant contributions appear below Africa and Indonesia. 
The strongest diffusion appears linked to intense patches of up/down-wellings in the equatorial belt at the CMB (see Figure~\ref{fig:R15_Maps_Diff}) and/or where strong gradient of ${\bf B}$ occur.
This is a direct consequence of our estimation of diffusion through cross-covariances involving core surface velocity and magnetic fields \citep[see BGA17 and][]{amit2008accounting}.
In the framework of our modelling, such diffusion patterns seem to be required by magnetic observations rather by the imposed prior cross-covariances (or if it is the case, it does not show up in the background state).

\subsection{Core flows solutions}
\label{sec:Core_flows}

Next, we study with more details the temporal information contained in our core flow solutions.
The idea is to extract an average signal and a linear acceleration, together with the flow at different periods, to check if we witness any preferential frequency, or if the characteristics of the flow change with the period.
To do so, we apply a least-squares regression to our core flow solution with a function of the form
\begin{linenomath*}\begin{eqnarray}
\label{eq:fit_LCS_U}
{\bf u}(t) = \hat{\bf A} + {\bf A}_L (t-t_0) + \displaystyle\sum_{k=1}^{11} \left[ {\bf A}_k^s \sin\left(2\pi(t-t_0)\dfrac{k}{T}\right) \right. \\
\left. + {\bf A}_k^c \cos\left(2\pi(t-t_0)\dfrac{k}{T}\right) \right]\,, \nonumber
\end{eqnarray}\end{linenomath*}
with $t_0=(t_i+t_f)/2=2008.92$ and $T = t_f-t_i=16$ yrs.
Vectors $\hat{\bf A}$, ${\bf A}_L$, ${\bf A}_k^c$ and ${\bf A}_k^s$ store respectively the spherical harmonic coefficients of the time average velocity, time average flow acceleration, and cosines and sines from periods $16$ yrs (for $k=1)$ to $1.45$ yrs (for $k=11$) -- of course the longer periods are not well constrained given the short time-span considered here.

\begin{figure*}
\centerline{
	\includegraphics[width=0.7\linewidth]{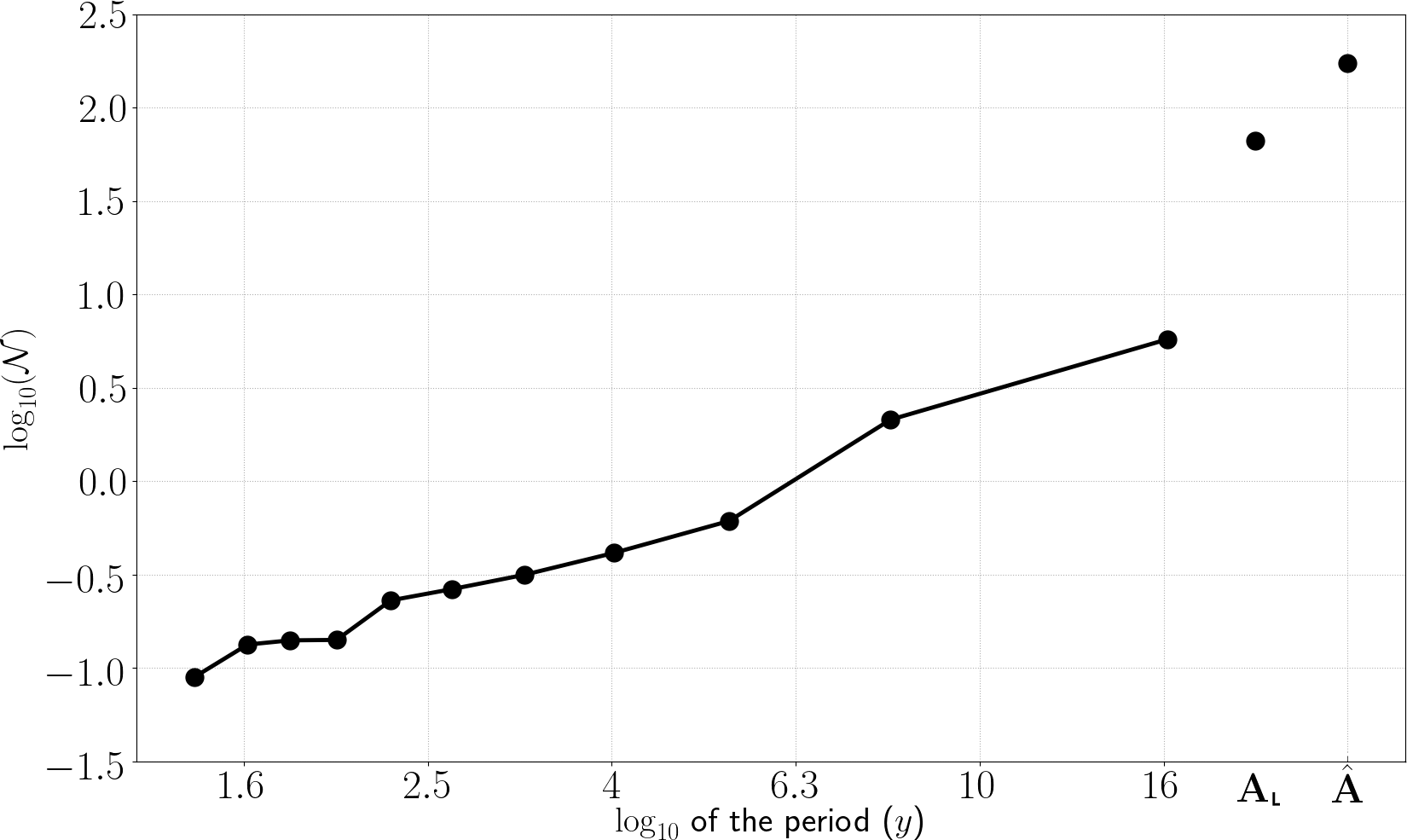}}
	\caption{
	Core flow norm ${\cal N}$ for all flow constituents that enter equation (\ref{eq:fit_LCS_U}). 
The norm $\mathcal{N}$ for the linear flow acceleration is obtained by integrating the linear trend over the 16 yrs.
	}
	\label{fig:R15_Fit_LCS_Spectrum}
\end{figure*}

We show in Figure~\ref{fig:R15_Fit_LCS_Spectrum} the norm (\ref{eq:norm_u}) of all flow constituents for the ensemble average solution. 
The flow is dominated by long periods, translating onto core surface motions the red SV temporal spectrum \citep[see][]{gillet2015stochastic,lesur2017frequency}.  
In comparison with a r.m.s. time average flow of 11.1 km/yr, the linear acceleration ${\bf A}_L$ corresponds, integrated over 16 yrs, to a r.m.s. flow increment of 6.6 km/yr. 

\subsubsection{Stationary motions, and flow model uncertainties}

\begin{figure*}
\centerline{
	\includegraphics[width=0.9\textwidth]{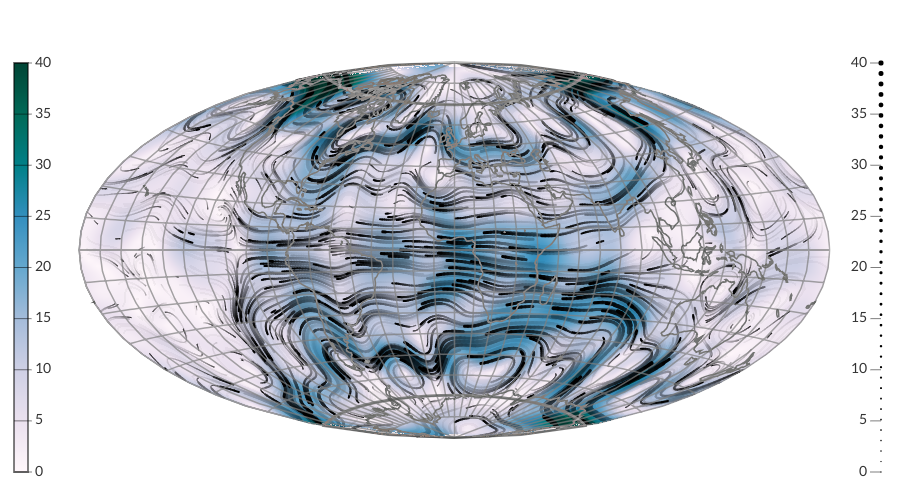}}
\centerline{
	\includegraphics[width=0.7\linewidth]{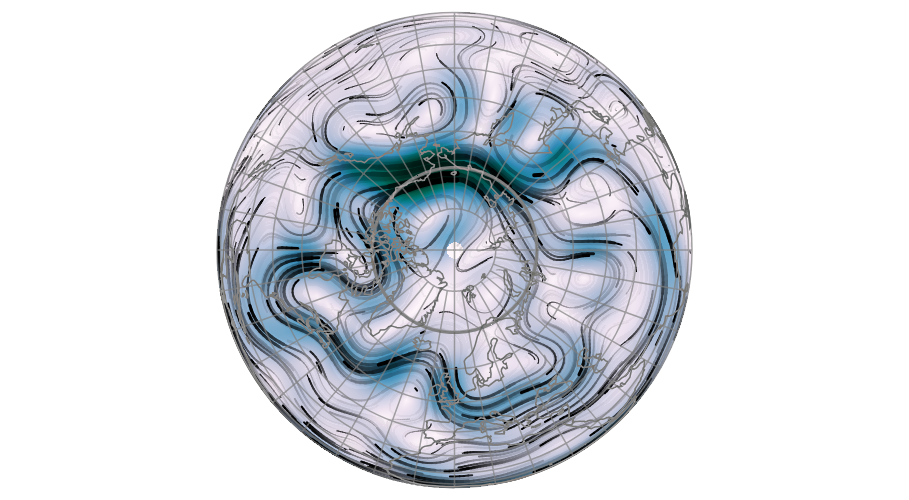}
	\hspace*{-2.5cm}
	\includegraphics[width=0.7\linewidth]{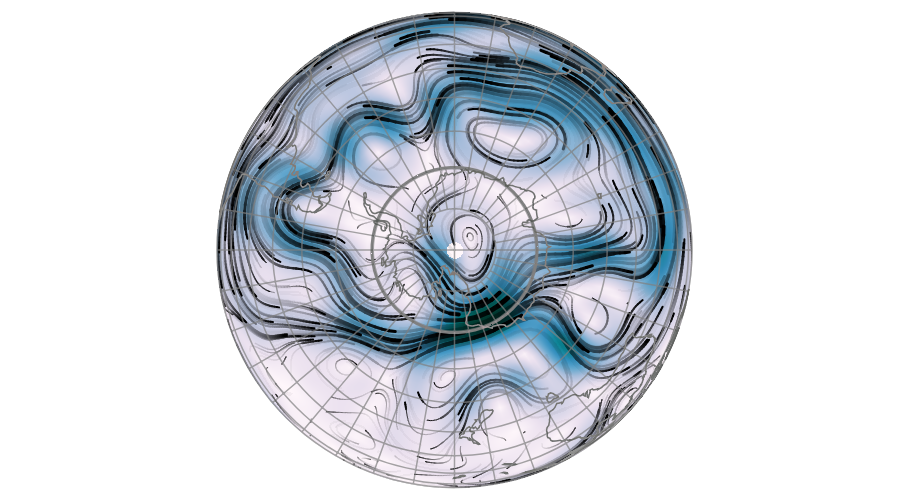}}
\centerline{
	\includegraphics[width=0.6\textwidth]{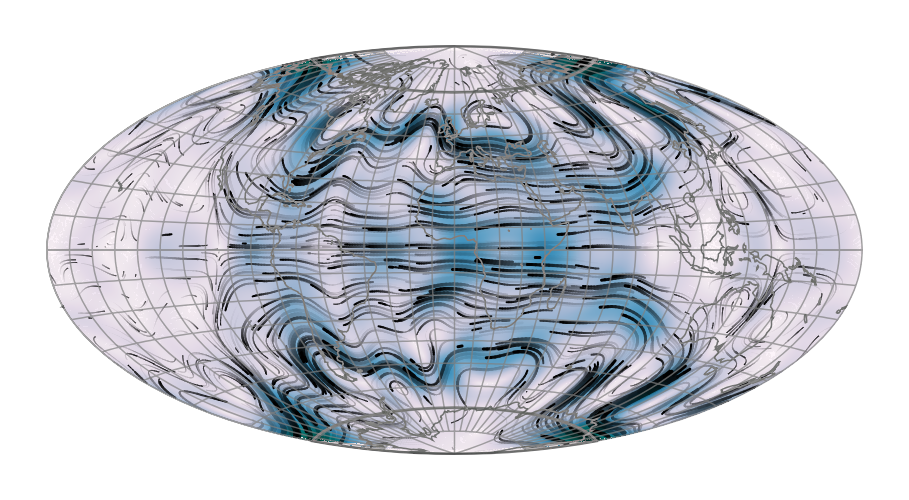}
	\hspace*{-.5cm}
	\includegraphics[width=0.6\textwidth]{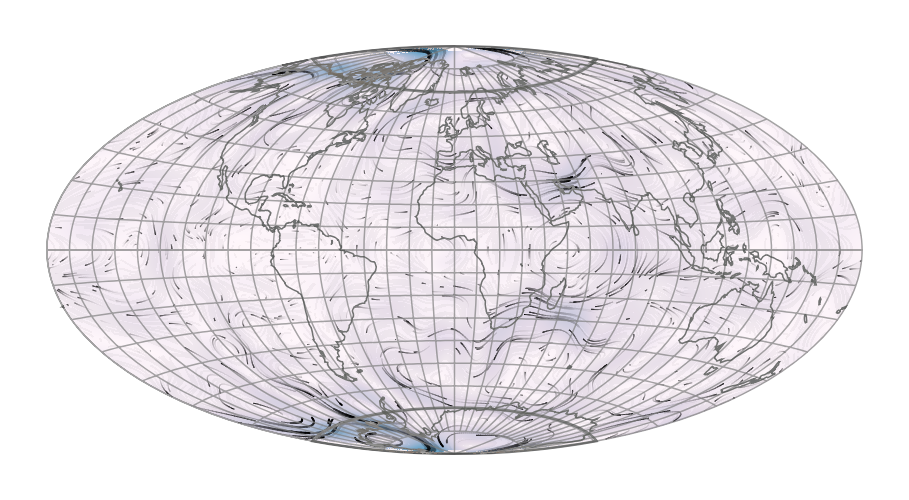}}
\caption{
Intensity maps at the CMB of the flow constituents $\hat{\bf A}$ (in km/yr) for the ensemble average flow solution, superimposed with passive tracers trajectories (black). 
Top: Aitoff projection. 
Middle: North (right) and South (left) polar projections.  
Bottom: Aitoff projection for equatorially symmetric (left) and antisymmetric (right) components. 
The colorscale and tracer size scale are the same for all sub-figures. 
	}
	\label{fig:R15_Fit_LCS_U stat}
\end{figure*}

We show in Figure \ref{fig:R15_Fit_LCS_U stat} core surface maps of the flow intensity and tracers trajectories for the ensemble average flow constituents $\hat{\bf A}$.
We retrieve on the map for the time average flow classical features, such as the westward gyre offset towards the Atlantic Ocean found in many studies \citep[e.g.][]{pais2008quasi,gillet2015planetary,aubert2014earth,baerenzung2017modeling}, with a Pacific hemisphere that is on average much less energetic.
The most energetic flow features are associated with (i) azimuthal motions in the equatorial belt below Africa, (ii)  high latitudes azimuthal jets in the Pacific hemisphere and (iii) meridional circulations, poleward (resp. equatorward) around 90$^{\circ}$W (resp. 90$^{\circ}$E).

Our solution is dominated by equatorially symmetric features (see Figure \ref{fig:R15_Fit_LCS_U stat}, bottom), as expected outside the tangent cylinder (or TC, the cylinder tangent to the inner core, whose axis coincides with to the rotation axis) when rotation forces dominates the momentum balance \citep[e.g.][]{pais2008quasi}. 
Nevertheless, the symmetry may be locally broken.
The most striking examples of this are anti-cyclonic circulations within the TC, retrieved in both the Northern and Southern hemispheres (Figure \ref{fig:R15_Fit_LCS_U stat}, middle).
In contrast with polar vortices previously inferred from geomagnetic observations \citep{olson1999polar,amit2006time}, features we isolate here are off-set to one side of the polar caps (i.e. they contain an important $m=1$ contribution). 
This is a common configuration for polar vortices found in the most up to date numerical simulations \citep{schaeffer2017turbulent}, which show much variability through epochs.

\begin{figure*}
\centering
	\includegraphics[width=.7\linewidth]{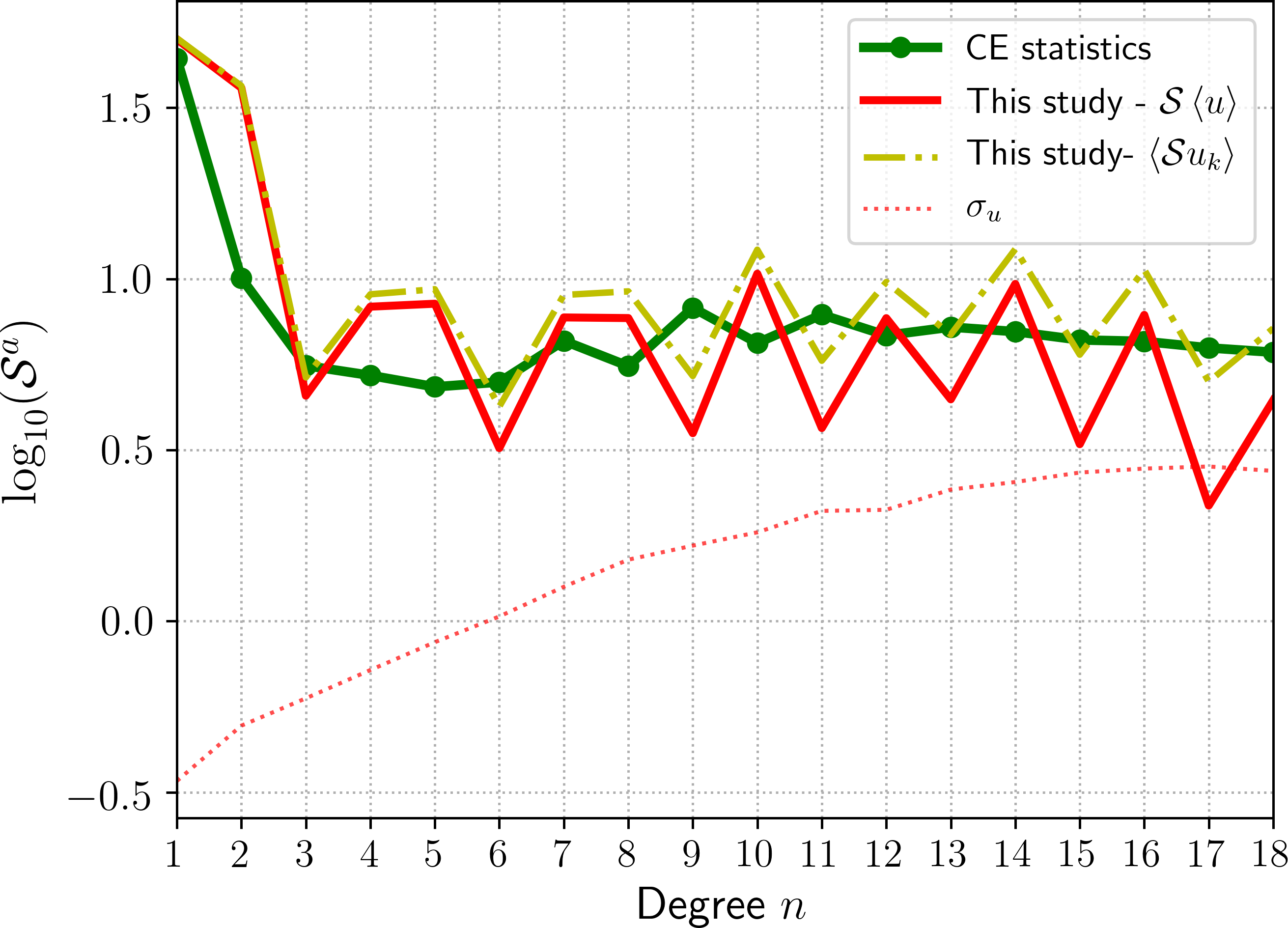}
	\caption{
	Time averaged spatial power spectra for ensemble average core flows ($\hat{\cal S}_{\left< u \right>}^a$, red thick line) and the spectra for the ensemble average of each realization ($\left< \hat{\cal S}_{uk}^a \right>$, yellow thick dotted line), eq. \ref{eq:spectrum_u_VO}): obtained from the re-analysis of VO and GO data. 
Spectra for the corresponding dispersion within the ensembles of models are displayed in dotted lines.
In green is shown the averaged spectrum for the prior CED.
	}
	\label{fig:R15_Spectra}
\end{figure*}

We show in Figure~\ref{fig:R15_Spectra} the time-average spatial power spectra for the ensemble average solution and for the dispersion within the ensemble of models.
The former is comparable with the spectrum of the prior CED.
The latter indicates that uncertainties, as measured by the ensemble spread, constitute a large fraction of the flow magnitude for degrees $n\ge 10$.
The oscillation in the power seen between odd and even degrees might be magnified by possibly too low subgrid error budget (see \S\ref{sec:Geomag_model}).

\subsubsection{On transient core surface motions}
\label{sec:SA_analysis}

\begin{figure*}
\centerline{
	\includegraphics[width=0.9\textwidth]{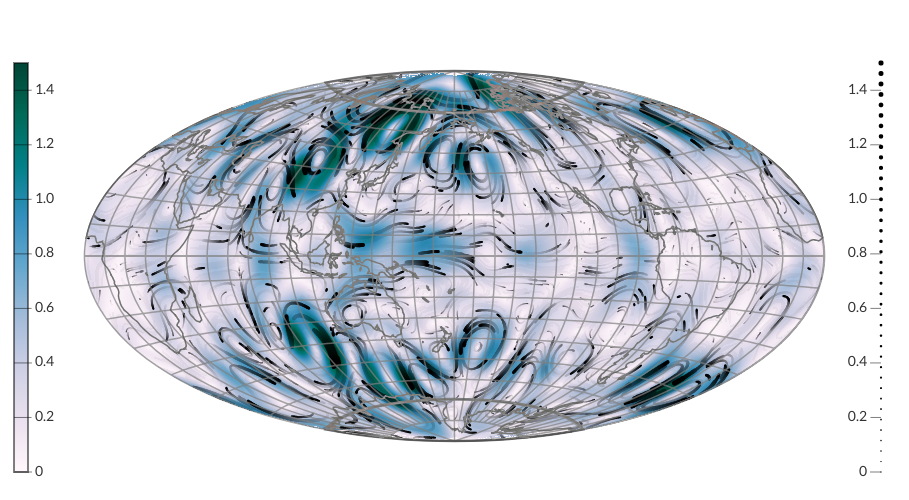}}
\centerline{
	\includegraphics[width=0.7\linewidth]{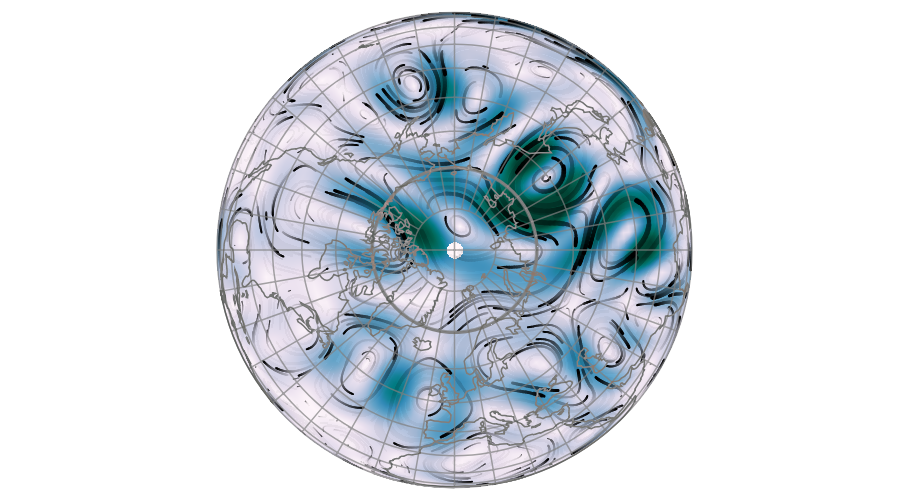}
	\hspace*{-2.5cm}
	\includegraphics[width=0.7\linewidth]{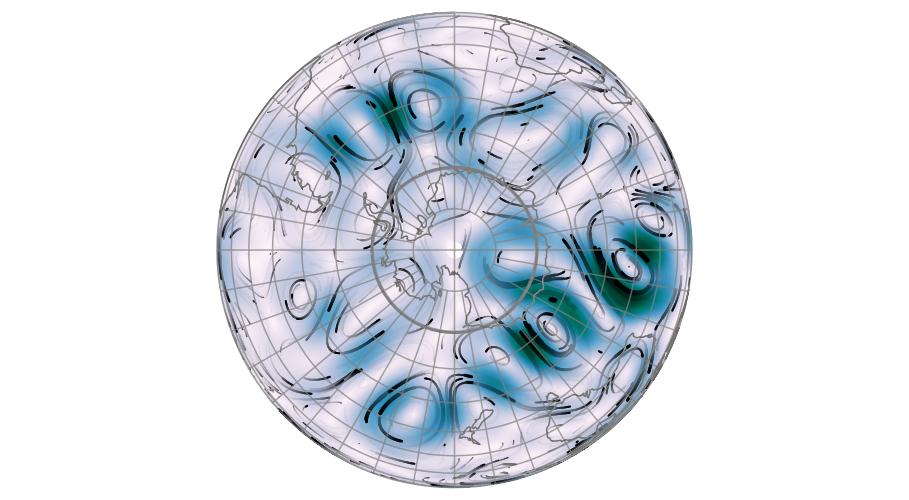}}
\centerline{
	\includegraphics[width=0.6\textwidth]{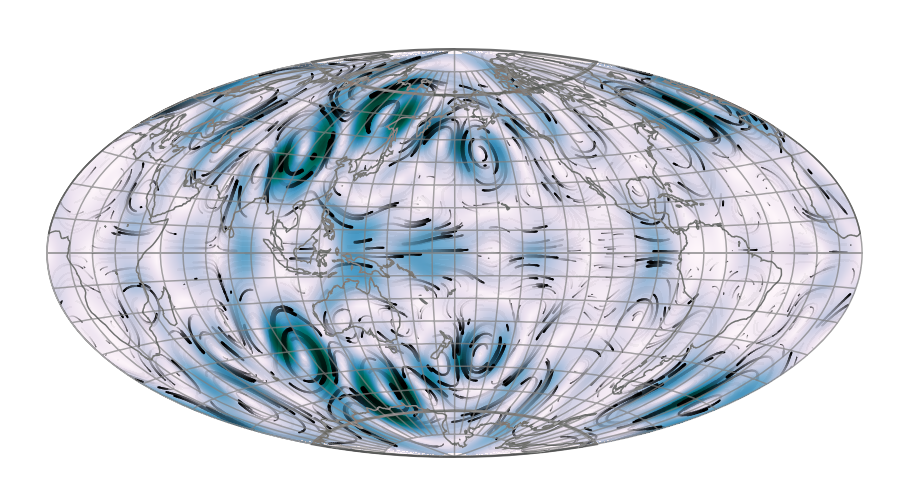}
	\hspace*{-.5cm}
	\includegraphics[width=0.6\textwidth]{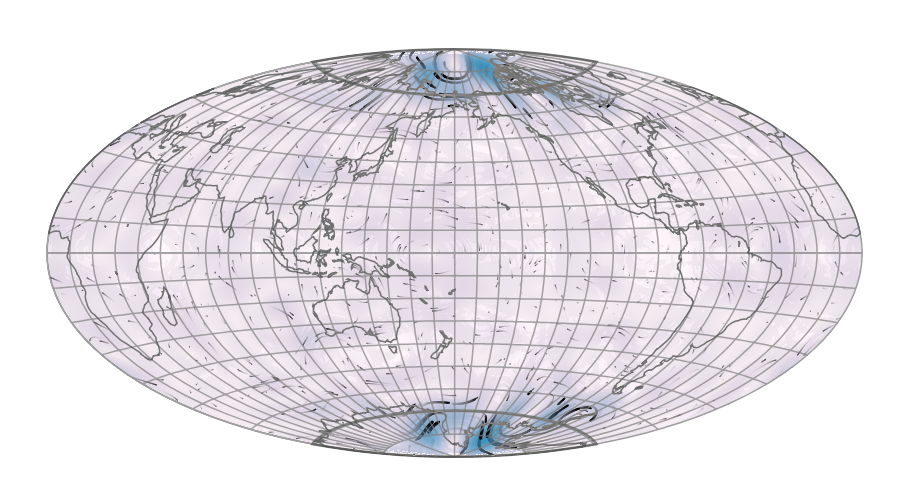}}
\caption{
Same as Figure \ref{fig:R15_Fit_LCS_U stat} for the flow constituent ${\bf A}_L$ (in km/yr$^2$).
	}
	\label{fig:R15_Fit_LCS_U lin}
\end{figure*}

We now explore transient flow motions. 
We particularly focus on the amount of equatorial symmetry of our solutions inside and outside the TC, in order to detect if our model is sensitive to the specific geometry of the Earth's core (does it hold a signature of the TC?).
As for the time-average flow, the linear acceleration over the past 16 yrs is primarily symmetric with respect to the equator (see Figure \ref{fig:R15_Fit_LCS_U lin}).  
The largest contributions consist of accelerating circulations around the meridional, Eastern branch of the gyre. 
Associated with these time-changing eddies around the equatorward branch of the planetary gyre, an Eastward equatorial jet intensifies under the Western Pacific.
This suggests an underlying dynamics more complex than a simple longitudinal shift of the planetary gyre.

Interestingly, our average solution does not show a major intensification of equatorially symmetric azimuthal jets at high latitudes in the Pacific hemisphere, as inferred by \cite{livermore2017accelerating}.
Indeed, we see an increase of the Northern jet only, by about $67\%$ in average (the one $\sigma$ dispersion within the ensemble of flow realizations allowing for an increase up to $100\%$). 
Although still an appreciable acceleration, it is significantly less than the factor of 3 found by \citeauthor{livermore2017accelerating}.
The disagreement is likely due to our global inversion (in opposition to their local model).
The difference seems to be related with anti-symmetric circulations within the TC.
One should keep in mind that in these high and low latitude areas, gradients of $B_r$ are much larger in the Northern hemisphere, meaning that the signature of any motions near the TC below the Southern Pacific are significantly weaker.
As for the stationary constituent, the equatorial symmetry is not perfectly respected, and we retrieve the largest anti-symmetrical features within the TC, associated with polar jets.

\begin{figure*}
\centerline{
	\includegraphics[width=0.9\textwidth]{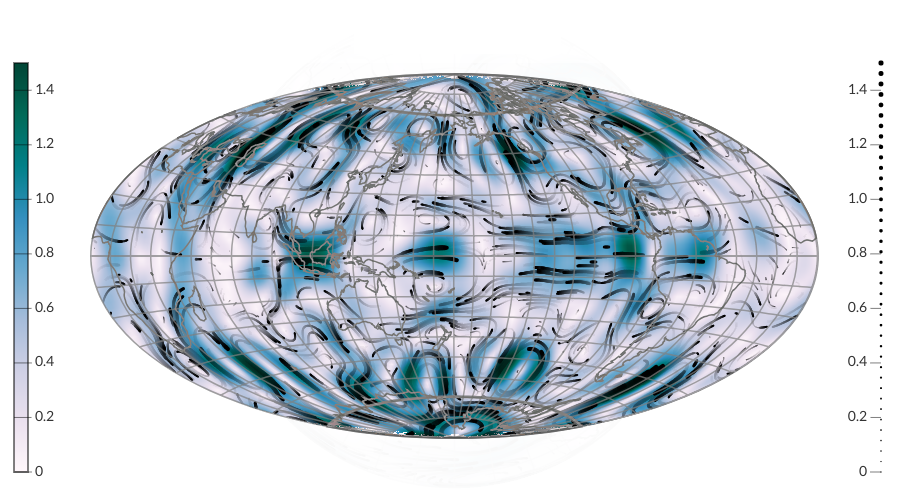}}
\centerline{
	\includegraphics[width=0.7\linewidth]{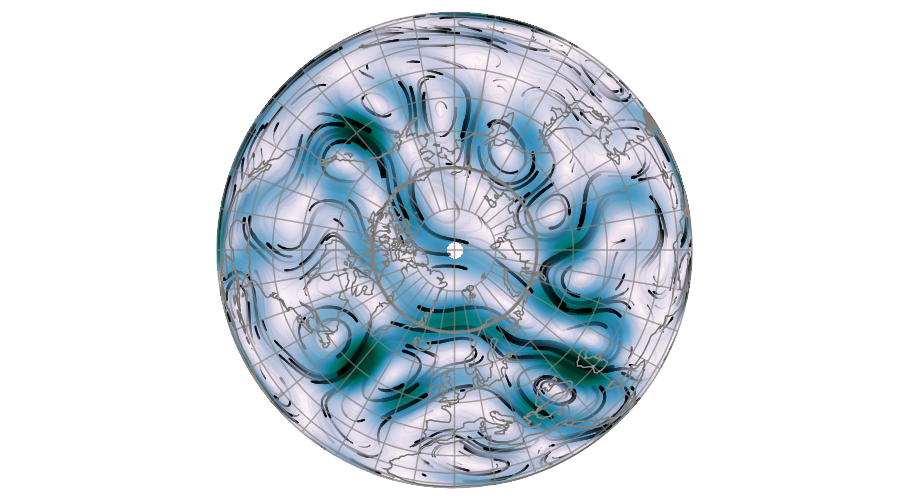}
	\hspace*{-2.5cm}
	\includegraphics[width=0.7\linewidth]{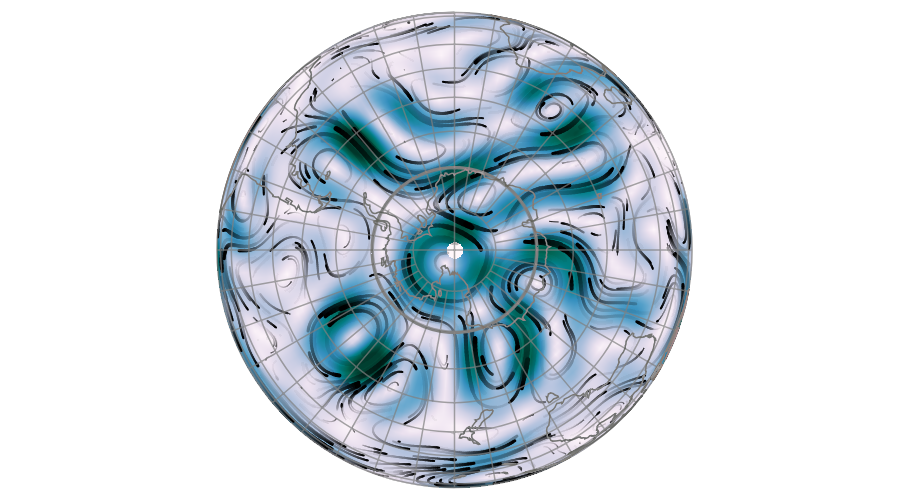}}
\centerline{
	\includegraphics[width=0.6\textwidth]{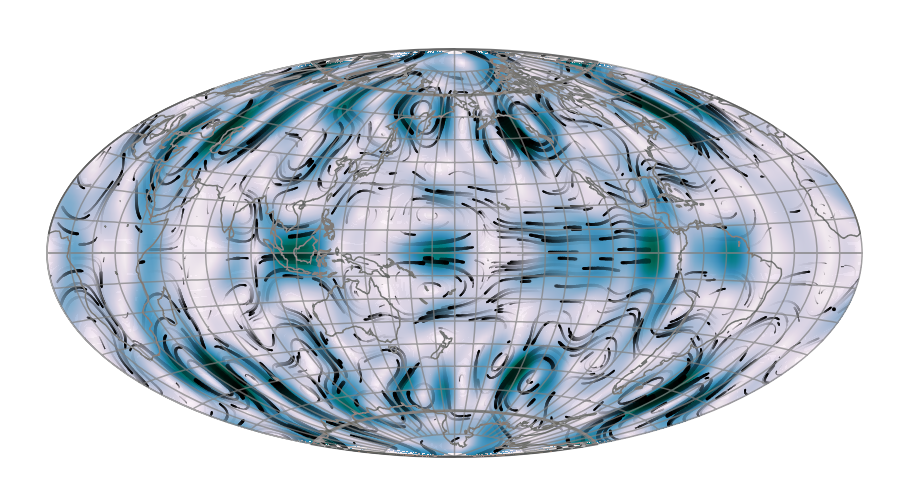}
	\hspace*{-.5cm}
	\includegraphics[width=0.6\textwidth]{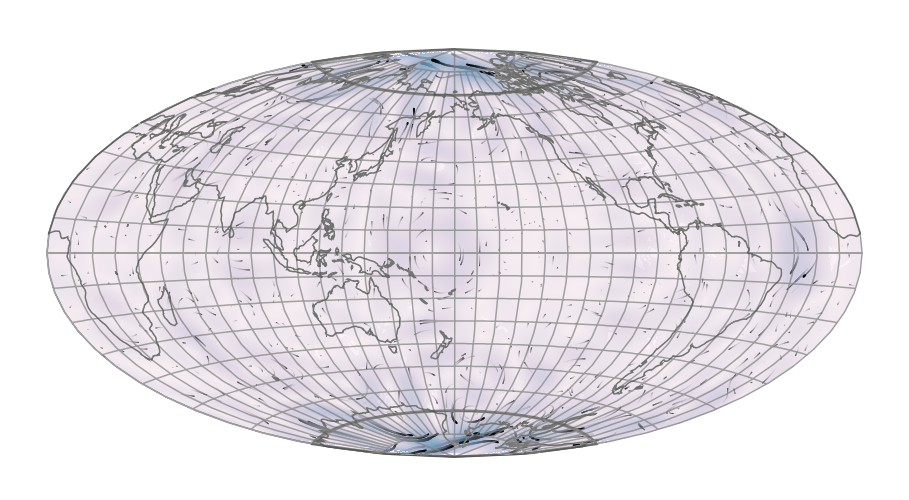}}
\caption{
Same as Figure \ref{fig:R15_Fit_LCS_U stat} for the flow constituent ${\bf A}_3^c$ (in km/yr).
	}
	\label{fig:R15_Fit_LCS_U cos3}
\end{figure*}

We give in Figure \ref{fig:R15_Fit_LCS_U cos3} an example of one interannual flow constituent at the CMB for a period of 5.3 yrs. 
In this case, the most energetic flows are concentrated into non-axisymmetric azimuthal jets near the equator \citep[already highlighted by][]{gillet2015planetary,finlay2016recent}, and into localised circulations at mid and high latitudes.
These are not confined to the Atlantic hemisphere: despite being less energetic on average, the Pacific hemisphere shows interesting interannual flow variations.
At these sub-decadal periods, we have not detected any obvious propagation of non-zonal flow patterns, which might be interpreted as the signature of azimuthally propagating waves \citep[as advocated for by][]{chulliat2014geomagnetic,chulliat2015fast}.
The other periods display globally the same kind of features and no particular behaviour is found at any period.
At these time-scales also show up less intense anti-symmetric features; the most significant shows up in the equatorial area (for instance under the Atlantic ocean and the Western Pacific), and towards high latitudes on the edge of the TC.

\begin{figure*}
\centerline{
	\includegraphics[width=0.7\linewidth]{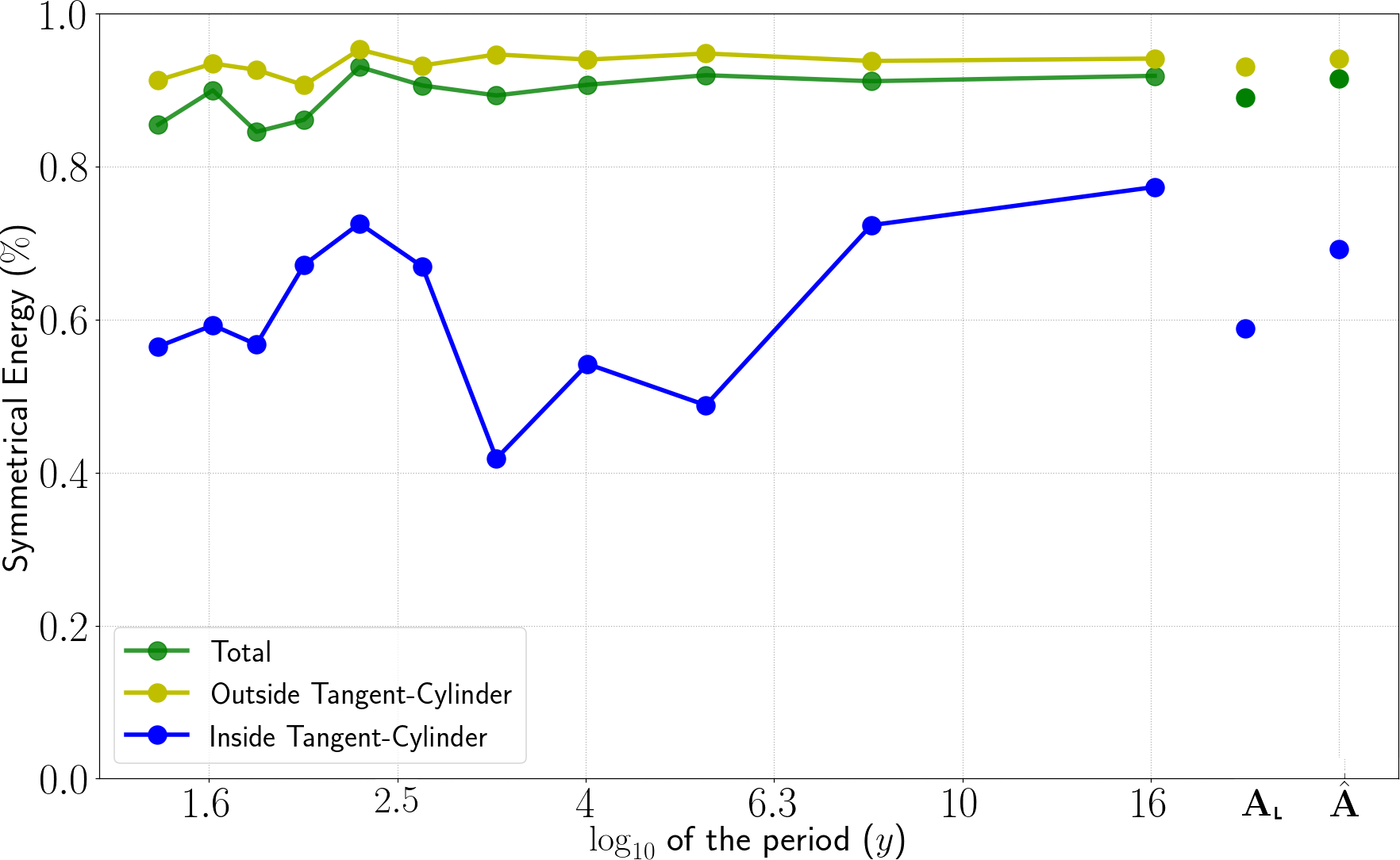}}
	\caption{
	Fraction of energy contained into the equatorial symmetric part of the flow, inside (blue line) and outside (yellow line) of the tangent cylinder (TC), for each of the flow constituent that enter equation (\ref{eq:fit_LCS_U}).
The total symmetric part of the flow is also displayed in green.
The value for the CED dynamo used as a prior, is $0.95$ both inside and outside the TC.
	}
	\label{fig:R15_Fit_LCS_Symmetries}
\end{figure*}

Figure~\ref{fig:R15_Fit_LCS_Symmetries} summarises the amount of equatorial symmetry found in regions inside and outside the TC, for our core flow solutions at all periods.
It appears almost independent of the considered period: 
outside the TC, it is within $90$ to $95\%$ of the surface energy for all flow constituents of equation (\ref{eq:fit_LCS_U}).
The partition of energy between symmetric and anti-symmetric flow components is more balanced inside the TC where, depending on the considered time-scale, $\approx 55\pm 15\%$ of the energy is contained in equatorially symmetric flows.
This latter observation could be expected because the presence of the inner core is intended to partially break the equatorial symmetry
However, it is remarkable that the algorithm appears accurate enough to detect a specific behaviour within the tiny areas covered by polar caps. 
Moreover, although our ensemble average model and the CED show very similar amounts of equatorial symmetry outside the TC (the value for the CED model is $95\%$ of symmetrical flows inside and outside TC), they differ significantly inside the TC (it is much less in the inverted flows).
As a consequence, the larger proportion of equatorial antisymmetry inside the TC is driven by observations (against the prior information).

\section{Summary and discussion}
\label{sec:Conclusion}

Following earlier strategies for geomagnetic field model reconstruction \citep[e.g.][]{jackson2007models,lesur2010modelling}, and moving towards geomagnetic data assimilation \citep{aubert2015geomagnetic,gillet2015stochastic,baerenzung2017modeling}, we continue the work initiated in BGA17.
We retain their idea of combining spatial information from numerical simulations of the geodynamo with temporal information implemented through stochastic equations, chosen to replicate the frequency spectrum of ground-based geomagnetic series.
However, instead of considering spherical harmonic coefficients of the main field as data, here we have inverted observations (GOs and VOs) directly, at and above the Earth's surface. 
In this respect we follow the studies by \citet{beggan2009forecasting} and \citet{whaler2015derivation}, although we account for subgrid processes \citep[of great importance, as shown by BGA17 or][]{baerenzung2016flow} and for surface magnetic diffusion.
This avenue allows us to propose PDFs for the main field and its secular variation, as well as for the recovered core motions.

\subsection{Geophysical insights}
\label{sec:conclusion_results}

The MF models presented here are consistent both with observations and with the imposed dynamical prior. 
The model uncertainties, as suggested by the ensemble spread, are slightly less than the distance of the average model to CHAOS-6. 
We recover in our core flow solutions a westward gyre that circulates around the TC at high latitudes in the Pacific hemisphere, and flows closer to the equator in the Atlantic hemisphere.
The largest contributions from magnetic diffusion are associated with up/down-wellings where the gyre meets the equatorial region (under Indonesia) and in the equatorial region below Africa.
At all time-scales, the flow is predominantly symmetric with respect to the equator, except inside the TC where the situation is more balanced (contrary to our dynamo prior that is mostly symmetric everywhere). 

The most intense time-average flow acceleration over the past 16 years is linked with evolving meanders around the equatorward branch of the gyre in the Eastern hemisphere, also associated with the appearance of an Eastward equatorial jet under the Western Pacific. 
We do find a decadal intensification of jets near the TC, although the magnitude of the acceleration we infer is lower than that estimated by \cite{livermore2017accelerating} with their reduced model. 
In our study, it is furthermore confined to the Northern hemisphere.
This equatorial asymmetry may be interpreted as the signature of an ageostrophic acceleration, keeping in mind that main field gradients are weak in the Southern Pacific, implying a weaker constraint on flow motions there \citep[see Figure 7 in][]{baerenzung2016flow}.
However, because our prior does not show any particular bias in those areas, it is likely that those features are mostly driven by the data.
On interannual periods, we find relatively energetic flow changes in both the Atlantic and the Pacific hemispheres, with both non-zonal equatorial jets and time-dependent mid-to-high latitudes eddies evident.

\subsection{Future work}
\label{sec:perspectives}

We currently lack a physical understanding for the features described above, whether it be through quasi-geostrophic flows \citep[e.g.][]{labbe2015magnetostrophic}, motions within a stratified layer \citep[e.g.][]{buffett2017}, or any other interpretation through a reduced model. 
We also lack suitable long coverage by high quality satellite records to perform spectral analyses with a refined sampling in the frequency domain, which would allow us to isolate possible waves at interannual periods.  
Development of such reduced models, and their coupling with stochastic processes for modelling unresolved processes, will be an important next step in our ability to understand and predict geomagnetic field changes. 

Meanwhile, our stochastic model itself could be improved; in particular it is desirable to avoid driving back the average trajectory towards an average dynamo simulation. 
This is indeed an unlikely state for the current era (say over decadal to centennial time-scales), which might be better represented by a re-analysis of for instance centennial motions from historical records \citep{jonkers2003four}.
Furthermore, because of the short time-span covered today by satellite data, we found it challenging to derive well-conditioned matrices for VO uncertainties.
This is a key-point for such data assimilation studies, which calls for further developments, e.g. through projections onto reduced basis in the data space.
Alternatively, we may wish to co-estimate, together with the core state, time-dependent external fields. 
Although possible, this calls for a severe re-encoding of both the forecast and analysis steps, in order to integrate satellite measurements along the tracks.

The general philosophy of our work is to retrieve information on the state of the Earth's core, and to provide realistic uncertainties on all state variables in a simple way.
The encouraging magnetic models obtained with this approach render our algorithm suitable for deriving candidates to the International Geomagnetic Reference Field \citep{thebault2015international}. 
Remaining in a stochastic framework, modifications of the forward model parametrisation -- such as accounting for a background state closer to the flow responsible for the magnetic field over the past decades -- may extend the prediction capability of our algorithm.
However, targeting accurate field predictions one will have to resort to deterministic (i.e. dynamically based) equations for the core state.

\section{Acknowledgements}

We thank Julien Aubert for providing the Coupled-Earth dynamo series used to build the core state statistics, and Loic Huder for spotting two errors in the code at the origin of the results of BGA17.
We also thank Julien Baerenzung and an anonymous referee for their useful comments that helped improve the quality of our manuscript.
We would like to thank as well GFZ German Research Centre for Geoscience for providing access to the CHAMP MAG-L3 data and to ESA for providing access to the Swarm L1b MAG-L data.
We also like to thank the staff of the geomagnetic observatories and INTERMAGNET for supplying high-quality observatory data.
NG and OB were partially supported by the French Centre National d'Etudes Spatiales (CNES) for the study of Earth's core dynamics in the context of the \textit{Swarm} mission of ESA. 
ISTerre is part of Labex OSUG@2020 (ANR10 LABX56), which with the CNES also finance the phd grant of OB. 
Numerical computations were performed at the Froggy platform of the CIMENT infrastructure (https://ciment.ujf-grenoble.fr) supported by the Rh\^one-Alpes region (GRANT CPER07 13 CIRA), the OSUG@2020 Labex (reference ANR10 LABX56) and the Equip@Meso project (referenceANR-10-EQPX-29-01).
MH and CF were supported by the Danish Council for Independent Research - Natural Sciences, Grant DFF-4002-00366. 

\bibliography{artbib}

\bibliographystyle{gji}

\label{lastpage}

\end{document}